\documentclass[12pt,a4paper,notitlepage]{article}

\usepackage{amsfonts}
\usepackage{amsmath}
\usepackage{amssymb}
\usepackage{amsthm}
\usepackage{color}
\usepackage[T1]{fontenc}
\usepackage[utf8]{inputenc}
\usepackage{graphicx}
\usepackage{hyperref}
\usepackage{mathrsfs}
\usepackage{multirow}
\usepackage[square]{natbib}
\usepackage[format=hang]{subcaption}
\usepackage{times}
\usepackage{upgreek}
\usepackage{bm}
\usepackage{bbm}
\usepackage{math}			
\usepackage{mathtools}
\usepackage{cleveref}
\AtBeginDocument{} 

\usepackage{pgf}
\usepackage{tikz}
\usepackage{pgfplots}
\pgfplotsset{compat=newest}
\usepackage{pgfplotstable}
\usepgfplotslibrary{groupplots}
\usepackage{color}
\usetikzlibrary{calc,backgrounds}  

\usepackage{import}

\usepackage[ruled,vlined,noend]{algorithm2e}
\SetKwInput{KwInput}{Input}      
\SetKwInput{KwOutput}{Output}
\SetKwInput{kwInit}{Initialize} 
\SetKwInput{kwOffline}{Offline} 
\SetKwInput{kwOnline}{Online}

\crefname{algocf}{alg.}{algs.}
\Crefname{algocf}{Algorithm}{Algorithms}

\usepackage{todonotes}

\definecolor{rwth1}{RGB}{0,84,159}      
\definecolor{rwth2}{RGB}{142,186,229}   
\definecolor{rwth3}{RGB}{0,97,101}      
\definecolor{rwth4}{RGB}{0,152,161}     
\definecolor{rwth5}{RGB}{87,171,39}     
\definecolor{rwth6}{RGB}{189,205,0}     
\definecolor{rwth7}{RGB}{255,237,0}     
\definecolor{rwth8}{RGB}{246,168,0}     
\definecolor{rwth9}{RGB}{227,0,102}     
\definecolor{rwth10}{RGB}{204,7,30}     
\definecolor{rwth11}{RGB}{161,16,53}    
\definecolor{rwth12}{RGB}{97,33,88}     
\definecolor{rwth13}{RGB}{122,111,172}  

\date{}

\graphicspath{{./images/}}

\begin{document}

\author{\large{Stephan Ritzert${}^{\,a,*}$, Jannick Kehls${}^{\,a}$,}\large{ Stefanie Reese${}^{\,a,b}$, Tim Brepols${}^{\,a}$ }\\[0.5cm] 
\hspace*{-0.1cm}
\normalsize{\em ${}^{a}$Institute of Applied Mechanics, RWTH Aachen
  University,}\\
\normalsize{\em Mies-van-der-Rohe-Str.\ 1, 52074 Aachen, Germany}\\
\normalsize{\em ${}^{b}$University of Siegen, 57076 Siegen, Germany}\\
\normalsize{\em \{stephan.ritzert, jannick.kehls, stefanie.reese, tim.brepols\}@ifam.rwth-aachen.de}\\[0.25cm]
}

\title{\LARGE Component based model order reduction with mortar tied contact for nonlinear quasi-static mechanical problems}
\maketitle

\small
{\bf Abstract.}
In this work, we present a model order reduction technique for nonlinear structures assembled from components. 
The reduced order model is constructed by reducing the substructures with proper orthogonal decomposition and connecting them by a mortar-tied contact formulation. 
The substructure projection matrices are computed by the proper orthogonal decomposition (POD) method from snapshots computed on the substructure level.
The snapshots are computed using Latin hypercube sampling based on a parametrization of the boundary conditions. 
In numerical examples, we show the accuracy and efficiency of the method for nonlinear problems involving material and geometric nonlinearities as well as non-matching meshes. 
The method can predict solutions of new systems with varying boundary conditions and material behaviors.

\vspace*{0.3cm}

\normalsize


\section{Introduction}

Many engineering structures are made of components. 
High-resolution finite element simulations of such systems are computationally expensive, especially for problems involving geometric and material nonlinearities. 
Model order reduction (MOR) can reduce this effort. 
In this work, we propose a method where the reduced order model (ROM) of a nonlinear modular system is constructed from ROMs of the substructures. 

The coupling of reduced substructures has a long history for linear dynamical systems. 
For an overview of the historical development of those methods, the reader is kindly referred to \cite{de_klerk_general_2008}.  
A very popular component mode synthesis method is the Craig-Bampton method (\cite{craig_coupling_1968}), where the internal degrees of freedom (DOFs) are approximated by a combination of eigenmodes and so-called static constraint modes. 
The static constraint modes linearly relate the boundary DOFs to the internal DOFs and are computed by a static condensation of the stiffness matrix. 
Not many works exist that extend these component mode synthesis methods to nonlinear problems. 
For geometric nonlinearities, there exist works by \cite{wenneker2013component}, \cite{kuether_modal_2016,kuether_modal_2017}, and \cite{bui_reduced-order_2024} that extend the Craig-Bampton method(\cite{craig_coupling_1968}).  
In \cite{kuether_modal_2016,kuether_modal_2017}, a Craig-Bampton approximation of the displacement is used where the internal DOFs are reduced eigenmodes and static constraint modes. 
All of these quantities are computed from the linear stiffness matrix. 
\cite{bui_reduced-order_2024} enhanced the Craig-Bampton approach by static modal derivatives to account for the nonlinearity.  
In all of these works, the nonlinear force vector and stiffness matrix are approximated by cubic polynomial functions.
This approximation is only valid for St. Venant-Kirchhoff materials and geometric nonlinearities.
For large strains and nonlinear materials, the static constraint modes would also be a function of the displacement and would need to be updated in every Newton iteration. 
These methods only hold for small strains with large rotations because the static constraint modes are computed from the linear stiffness matrix.

In this work, we propose a data-driven reduction method where the DOFs of each substructure are reduced by the proper orthogonal decomposition (POD) method. 
The POD-Galerkin method (\cite{lenaerts2001proper}) reduces the size of the system of equations and increases the speed of the solution of the solution. 
In the past it was successfully applied to solve nonlinear solid mechanics problems, see e.g. \cite{lenaerts2001proper,herkt2009model,radermacher2013comparison,radermacher_proper_2013}. 
An overview of POD in the field of structural dynamics is given in \cite{kerschen_method_2005}.  
A drawback of POD based model order reduction is that the solution time still depends on the original size of the problem, because in every iteration the nonlinear force vector and the tangential stiffness matrix have to be assembled. 
Therefore, different hyperreduction techniques where developed that can reduce the computational time to assemble the system. 
For recent works of POD based hyperreduction techniques in nonlinear mechanics, 
see among others: energy conserving sampling and weighting (\cite{farhat_structure-preserving_2015}),  discrete empirical interpolation method for nonlinear solid mechanics (\cite{radermacher_pod-based_2016,ghavamian2017pod}),  (continuous) empirical cubature method (\cite{hernandez_dimensional_2017,hernandez_cecm_2024}),  hyperreduction for nonlinear structural dynamics \cite{rutzmoser_model_2018},  hyperreduction for nonlinear computational homogenization problems (e.g. \cite{guo2024reduced,wulfinghoff2024statistically}). 
In this work, we present a POD-based component-wise MOR method, which is the foundation of future works including hyperreduction.

Not many works use POD-based MOR in substructuring problems, where the ROM of the whole system is assembled from the ROMs of the components. 
For linear problems component-wise POD was used in \cite{ritzert_adaptive_2023}, where it was applied to parametric substructures and in \cite{mcbane_stress-constrained_2022}, where it was used for topology optimization of lattice-like structures. 
In nonlinear mechanics \cite{zhou_proper_2018} proposed a method where the components are reduced by POD and connected by a penalty method. 
The penalty matrices are also reduced by parts of the substructure POD bases. 
In \cite{hernandez_multiscale_2020} a method was proposed to compute periodic structures by another component-based MOR approach. 
In this approach, the substructures are coupled by fictitious interfaces. 

In this work, we propose a new approach. 
We use a mortar-tied contact formulation as a basis and reduce the degrees of freedom of the substructures by individual pre-computed POD bases. 
The mortar method is a state-of-the-art contact mechanics method for non-matching meshes. 
An advantage of the method is, that the Lagrange multipliers, as well as the slave-side DOFs, can be removed from the equation system by static condensation, by choosing dual shape functions for the Lagrange multipliers.
This property is used in this MOR technique.   
For more information regarding the mortar method for tied-contact, the reader is referred to e.g. \cite{wohlmuth_discretization_2001,puso_3d_2004,laursen_mortar_2012,popp_contact_2018}.
Each substructure is reduced by POD projection matrices.
The slave side interface DOFs are not reduced because they can be related to the reduced interface displacements of the master side.

In the numerical examples, we discuss the performance and accuracy of our method. 
We investigate large deformations, stability effects, different stiffnesses of the substructures, non-matching meshes and viscoelastic material behavior. 
For all numerical examples, the snapshots are computed on the substructure level. 
We parametrize the displacement boundary conditions and use a Latin hypercube sampling (LHS) procedure. 

\paragraph{Outline of the paper} 
In \Cref{sec:equations}, we first discuss the mortar tied-contact full order model (\Cref{sec:FOM}) and then explain the component-based model order reduction technique, where the substructures are first reduced and then connected by the reduced mortar tied-contact (\Cref{sec:ROM}).  
The accuracy and performance of the MOR technique and the snapshot sampling are investigated in the numerical examples in \Cref{sec:examples}.  
Finally, the results are discussed in \Cref{sec:conclusion}.

\section{Component based model order reduction}\label{sec:equations} 

\subsection{Full-order model}\label{sec:FOM} 

The full-order model (FOM) from which we derive our reduced-order model (ROM) is a mortar-tied-contact formulation with static condensation of the Lagrange multipliers and the contact displacements on the slave side. 
In the following, it is briefly derived.  

\subsubsection*{Weak formulation} 
\begin{figure}[h!]
\centering
\begin{tikzpicture}
    \node[inner sep=0pt] (pic) at (0,0) {\includegraphics[width=0.35\textwidth]
        {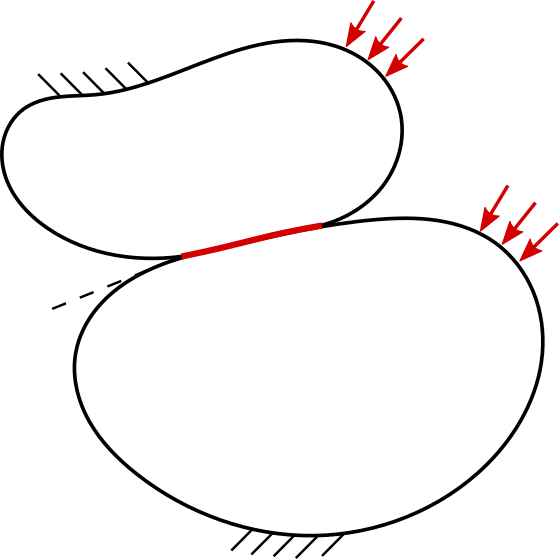} };
    \node[inner sep=0pt] (b1)   at ($(pic.center)+(-1.7,0.7)$)  {$\Omega_0^1$  };
    \node[inner sep=0pt] (b2)   at ($(pic.center)+(0.5,-1.0)$)  {$\Omega_0^2$  };
    \node[inner sep=0pt] (b2)   at ($(pic.center)+(-0.2,0.75)$)  {$\Gamma_c^{1,2}$  };
    \node[inner sep=0pt] (b2)   at ($(pic.center)+(0.20,-2.1)$)  {$\Gamma_u^{2}$  };
    \node[inner sep=0pt] (b2)   at ($(pic.center)+(-1.20,1.6)$)  {$\Gamma_u^{1}$  };
    \node[inner sep=0pt] (b2)   at ($(pic.center)+(0.5,1.85)$)  {$\Gamma_\sigma^{1}$  };
    \node[inner sep=0pt] (b2)   at ($(pic.center)+(2,-0.0)$)  {$\Gamma_\sigma^{2}$  };
    \node[inner sep=0pt] (b2)   at ($(pic.center)+(-3.2,0.0)$)  {$\bu_c^1 = \bu_c^2 $  };
    \node[inner sep=0pt] (b2)   at ($(pic.center)+(-3.3,-0.5)$)  { $\bt^1 = - \bt^2 $}; 
\end{tikzpicture}
\caption{Illustration of a domain, that is composed of two subdomains $\Omega_0^1$ and $\Omega_0^2$, with the Neumann boundaries $\Gamma_\sigma^1$, and $\Gamma_\sigma^2$ and Dirichlet boundaries $\Gamma_u^1$, and$\Gamma_u^2$. At the contact interface $\Gamma_c^{1,2}$, the two tied contact conditions are displayed. } 
\label{fig:tiedContactKartoffeln}
\end{figure}

In tied-contact problems, we introduce two conditions that must be fulfilled in addition to the balance of linear momentum.  
The first condition is that the displacements of two bodies must be equal on the contact interface $\Gamma_c^{1,2}$:
\begin{equation}
    \bu_c^1 = \bu_c^2 \quad \text{on } \Gamma_c^{1,2}
\end{equation}
The second condition is the equilibrium of the traction acting on the two bodies at the contact interface:
\begin{equation}
    -\bt_c^1 = \bt_c^2 \quad \text{on } \Gamma_c^{1,2}
\end{equation}
A tied-contact problem consisting of two substructures is displayed in \Cref*{fig:tiedContactKartoffeln}.

In this work, we use the Lagrange multiplier method to enforce those conditions. 
We introduce the Lagrange multiplier $\Blambda = - \bt_c^1 = \bt_c^2$ and come up with the variational saddle-point problem that can be derived from the Lagrange functional (see e.g. \cite{wohlmuth_discretization_2001}):
\begin{equation}\label{eq:saddlePoint} 
\begin{aligned}
    \sum_{i=1}^{n_s} \delta g^i_{\rm int}(\bu,\delta\bu) + \sum_{j=1}^{n_c} \delta g^j_{c,u}(\Blambda,\delta\bu_c^1,\delta \bu_c^2) &= 0 \\
    \sum_{j=1}^{n_c} \delta g^j_{c,\lambda} ( \bu_c^1,\bu_c^2,\delta \Blambda) &= 0 
\end{aligned}
\end{equation} 
Here $n_s$ is the number of substructures and $n_c$ is the number of tied-contact interfaces. 
The displacements at the tied contact interfaces denoted by $\bu_c^1,\bu_c^2 $ are subsets of the displacements of the respective substructures that share that interface.  
The weak form of the balance of linear momentum for the quasi-static case is 
\begin{equation}\label{eq:internal_energy}
    \delta g_{\rm int}^i = \int_{\Omega_0} \left( \bS : \delta \bE \, - \rho \bb \cdot \delta \bu \right) \di V  - \int_{\Gamma_t} \delta \bu \cdot \bm{t} \, \di A,
\end{equation}
where $\bS$ is the second Piola-Kirchhoff stress tensor, $\delta \bE$ is the virtual Green-Lagrange strain tensor, $\bb$ are the body forces, $\rho$ is the density and $\bt$ are the tractions acting on the boundary $\Gamma_t$. 
From the variational Lagrange multiplier method, we get the following contributions that enforce the tied-contact constraints 
\begin{align}
\label{eq:weakcontact0}
    \delta g^j_{c,u} &=  \int_{\Gamma_c^{1,2}}  \left( \delta \bu_c^1 - \delta \bu_c^2 \right) \cdot \, \Blambda \, \di A \\ 
    \delta g^j_{c,\lambda} &=  \int_{\Gamma_c^{1,2}}  \delta \Blambda \cdot \left( \bu_c^1 -  \bu_c^2 \right) \, \di A
\label{eq:weakcontact}
\end{align}

\subsubsection*{Discretization}
To solve \Cref{eq:saddlePoint}, we discretize \Cref{eq:internal_energy} to \Cref{eq:weakcontact} with the finite element method (FEM). 
A displacement-based discretization of \Cref{eq:internal_energy} leads to the nonlinear $n \times 1$-dimensional vector equation 
\begin{equation}
    \bG(\bU(t),t) = \bR(\bU(t)) - \bF_{\rm ext}(t).
\end{equation}
$\bR(\bU(t))$ is the internal force vector, that depends nonlinearly on the displacement vector $\bR(\bU(t))  $ at time $t$, and $\bF_{\rm ext}(t)$ is the external force vector.
The discretization of \Cref{eq:weakcontact0} and \Cref{eq:weakcontact} yields 
\begin{align}
    \delta g_{c,u} &= \delta (\bU_c^1)^T \bD^T \BLambda - \delta(\bU_c^2)^T \bM^T \BLambda \\
    \delta g_{c,\lambda} &= \delta \BLambda^T \bD \bU_c^1 - \delta  \BLambda^T \bM \bU_c^2
\end{align}
where we introduce the mortar-matrices $\bD$ and $\bM$. 
These matrices arise from a finite element discretization of the interface displacements $\bu_c^i = \bN^i \,  \bU_c^i $ and the Lagrange multipliers $\Blambda = \hat \bN^1 \BLambda$. 
$\bN^1$ and $\bN^2$ are shape functions for the displacements and the coordinates of the slave and master side interface elements.
$\hat \bN^1$ are the shape functions of the Lagrange multiplier, defined on the slave side of the interface.  
The mortar matrices are computed for all $n_e$ interface elements and assembled into the matrices $\bD, \bM$. They are defined as: 
\begin{align}
    \bD = \Ass_{e=1}^{n_e}\,  \bD^e,  \qquad \bD^e = \int_{\Gamma_e^1} (\hat\bN^1)^T \bN^1 \, \di A \\ 
    \bM = \Ass_{e=1}^{n_e}\,  \bM^e,  \qquad \bM^e = \int_{\Gamma_e^1} (\hat\bN^1)^T \bN^2 \, \di A.
\end{align} 
Here, $\Ass$ is the assembly operator.
The integrals on the element level can be solved by the mortar method (e.g. \cite{puso_3d_2004,popp_contact_2018}). 
The advantage of this method is, that the integrals can be solved for non-conforming meshes. 

The discretized equations for a system consisting of $n_s$ substructures and 
$n_c$ tied-contact interfaces reads 
\begin{align}
    \label{eq:fullFEMequation0}
    \sum_{i=1}^{n_s} \bG_i(t,\bU_i)  +  \sum_{j=1}^{n_c} (\bD^T_j \BLambda_j - \bM^T_j \BLambda_j) &= 0 \\ 
     \sum_{j=1}^{n_c} (\bD_j \bU_{c,j}^1 - \bM_j \bU_{c,j}^2) &= 0 
     \label{eq:fullFEMequation1} 
\end{align}
Note, that the displacements of each contact interface $j$: $\bU_{c,j}^1$ and $\bU_{c,j}^2$ are a subset of the displacements of a substructure $\bU_{i}$.
Since $\bG_i(t,\bU_i)$ is a nonlinear vector, the above system of equations (\Cref{eq:fullFEMequation0,eq:fullFEMequation1}) will be solved by the Newton-Raphson scheme. 
For the linearization of \Cref{eq:fullFEMequation0} we define the tangential stiffness matrix of a substructure $i$ for the displacement state $\bU_i^k$ as $\bK_i(\bU_i^k) \coloneqq \frac{\partial \bG_i}{\partial \bU_i} |_{\bU_i^k}$. 
To compute the solution of the $k+1$ iteration we solve the following equation system for all $\Delta \bU_i$
\begin{align}\label{eq:NewtonRaphson0}
    \sum_{i=1}^{n_s} \left( \bG_i(t,\bU_i^{k}) + \bK_i(\bU_i^{k}) \, \Delta \bU_i \right)+ \sum_{j=1}^{n_c} \left(\bD^T_j \BLambda_j^k - \bM^T_j \BLambda_j^k \right) &= \bm 0 \\ 
    \sum_{j=1}^{n_c} \left(\bD_j \, \Delta\bU_{c,j}^1 - \bM_j \, \Delta\bU_{c,j}^2 \right) &= \bm 0 
    \label{eq:NewtonRaphson1}
\end{align}
The displacements of all substructures are then updated by $\bU_i^{k+1} = \bU_i^{k} + \Delta \bU_i$. 
This procedure is repeated until \Cref{eq:fullFEMequation0,eq:fullFEMequation1} are approximately fulfilled. 

For simplicity, further derivations are shown for a system with two substructures, where one body is called master and the other is called slave. 
From now on, the master degrees of freedom will be denoted by a superscript $M$ and the slave degrees of freedom by a superscript $S$. 
The displacements and residual vectors of each substructure are split into internal degrees of freedom $\bU_I, \, \bG_I$ and tied contact degrees of freedom $\bU_C, \, \bG_C$.  
The tangential stiffness matrices are split accordingly into four block matrices. 
For this special case, \Cref{eq:NewtonRaphson0,eq:NewtonRaphson1} can be written in matrix-vector notation:
\begin{equation}\label{eq:firstLGS}
    \begin{bmatrix}
\bK_{II}^M & \bm{0} & \bK_{IC}^M & \bm{0} & \bm{0} \\ 
\bm{0} & \bK_{II}^S & \bm{0} & \bK_{IC}^S & \bm{0} \\ 
\bK_{CI}^M & \bm{0} & \bK_{CC}^M & \bm{0} & - \bM^T \\ 
\bm{0} & \bK_{CI}^S & \bm{0} & \bK_{CC}^S & \bD^T \\ 
\bm{0} & \bm{0} & -\bM & \bD & \bm{0}
\end{bmatrix}
\begin{bmatrix} 
\Delta \bU_I^M \\ \Delta \bU_I^S \\ \Delta \bU_C^M\\ \Delta \bU_C^S \\ \BLambda
\end{bmatrix} = -
\begin{bmatrix} 
\bG_I^M \\ \bG_I^S \\ \bG_C^M \\ \bG_C^S \\ \bm 0 
\end{bmatrix} 
\end{equation}
This equation system has the typical saddle point structure with a zero block matrix on the main diagonal. 
It can have both positive and negative eigenvalues and is therefore indefinite(cf. e.g. \cite{rusten1992preconditioned}).
The saddle point structure is unfavorable for efficient solution algorithms. 
It requires special preconditioning techniques to apply iterative solution methods like the conjugate gradient (CG), or the generalized minimal residual (GMRES) method (\cite{saad2003iterative,benzi2005numerical}).   
Moreover, the additional degrees of freedom would require additional mode matrices for the ROM. 
In this paper, we use the mortar method to compute the matrices $\bD$ and $\bM$ (\cite{wohlmuth_discretization_2001,laursen_mortar_2012,popp_contact_2018}). 
The advantage of this method is, that the Lagrange multipliers can be removed from the system of equations in \Cref{eq:firstLGS}, and it can be transformed into a symmetric positive definite system of equations. 
Another advantage is the integration procedure to compute the mortar matrices $\bD$ and $\bM$, which allows for non-conforming meshes. 
The integration procedure is explained in detail in e.g. \cite{puso_3d_2004, popp_contact_2018}.


\subsubsection*{Static condensation}
In this section, we describe how the system of equations in \Cref{eq:firstLGS} can be transformed into a symmetric positive definite system by static condensation.  
From the fifth equation in \Cref{eq:firstLGS}, it follows that: 
\begin{equation}\label{eq:interfaceCoupling}
    \Delta \bU_C^S = \bD^{-1}\bM \, \Delta \bU_C^M \coloneqq \bP \, \Delta \bU_C^M
\end{equation} 
where $\bP \coloneqq \bD^{-1}\bM$ is the discrete interface coupling operator (\cite{popp_contact_2018}). 
The Lagrange multipliers $\BLambda$ can be expressed as 
\begin{equation}\label{eq:disLagrangeMult}
    \BLambda = \bD^{-T} \left( -\bg_c^S - \bK_{CI}^S \, \Delta \bU_I^S - \bK_{CC}^S \, \Delta \bU_C^s \right)
\end{equation} 
by reordering the fourth equation of the equation system given in \Cref{eq:firstLGS}. 
By inserting \Cref{eq:disLagrangeMult} into the second and third equation of \Cref{eq:firstLGS} and utilizing \Cref{eq:interfaceCoupling} we obtain the final condensed system of equations: 
\begin{equation}\label{eq:condensedLGS}
    \underbrace{
\begin{bmatrix}
\bK_{II}^M & \bm{0} & \bK_{IC}^M  \\ 
\bm{0} & \bK_{II}^S & \bK_{IC}^S \, \bP \\ 
\bK_{CI}^M & \bP^T \bK_{CI}^S & \bK_{CC}^M + \bP^T \bK_{CC}^S \bP  \\ 
\end{bmatrix}}_{\bK_{\rm cond}}
\underbrace{\begin{bmatrix} 
\Delta \bU_I^M \\ \Delta \bU_I^S \\ \Delta \bU_C^M
\end{bmatrix}}_{\Delta \bU_{\rm cond}} = -
\underbrace{\begin{bmatrix} 
\bG_I^M \\ \bG_I^S \\ \bG_C^M + \bP^T \bG_C^S
\end{bmatrix}}_{\bG_{\rm cond}}
\end{equation}
Here also the definition of the interface coupling operator $\bP \coloneqq \bD^{-1}\bM$ is used. 
This static condensation requires the inversion of the mortar matrix $\bD$, which for standard shape functions is computationally expensive.
A solution to this problem was proposed by \cite{wohlmuth_discretization_2001} by choosing dual shape functions for the Lagrange multipliers. 
Because of these dual shape functions, the mortar matrix $\bD$ reduces to a diagonal matrix that can be easily inverted.

The dual Lagrange multiplier shape functions $\hat N_j^S$ are defined by the biorthogonality condition with the displacement shape functions $N_k^{S}$: 
\begin{equation}\label{eq:biorthogonality}
    \int_{\Gamma_e} \hat N_j^S N_k^S \, \di A = \delta_{jk} \int_{\Gamma_e} N_k^S \, \di A
\end{equation}
which was first presented in the works of \cite{scott1990finite} and applied to the mortar method in \cite{wohlmuth_discretization_2001}.
The Lagrange multiplier shape functions are computed by a linear mapping of the standard displacement shape functions 
\begin{equation}
    \hat N_j^S = a_{jk} N_k^S
\end{equation}
with the coefficient matrix $\bA = [a_{jk}]$. 
The coefficients are obtained by the expression 
\begin{equation}
    \bA = \bB \bC^{-1},
\end{equation} 
stemming from the biorthogonality condition \Cref{eq:biorthogonality}.
The entries of the matrices $\bB = [b_{jk}]$ and $\bC = [c_{jk}]$ are computed by 
\begin{align}
     b_{jk} &= \delta_{jk} \int_{\Gamma_e} N_k \, \di A \\ 
      c_{jk} &= \int_{\Gamma_e} N_j N_k \, \di A. 
\end{align}
    
With the method described above, the condensed system of equations can be efficiently computed. 
It should be noted, that the interface-coupling operator $\bP$ only needs to be computed in the beginning. 
Only the residual vectors and tangential stiffness matrices of the substructures have to be evaluated in every iteration step. 
The condensed system of equations \Cref{eq:condensedLGS} is the basis of the reduced-order model described in the following section.

\subsection{Reduced-order model}\label{sec:ROM}

\subsubsection*{Component based model order reduction} \label{sec:componentROM}
The idea of this model order reduction (MOR) technique is to reduce the DOFs of the substructures on their own and assemble them into a global reduced system of equations. 
The difference to the MOR techniques in \cite{zhou_proper_2018,ritzert_adaptive_2023} is that we here use the mortar method for mesh-tying (cf. \Cref{sec:FOM}) instead of the penalty approach. 

For the derivation of the ROM, we use a system composed of two substructures denoted by the superscripts $S$ for the slave side and $M$ for the master side (cf. \Cref{eq:firstLGS} to \Cref{eq:condensedLGS}).
The displacement vectors of the substructures are split into internal displacements $\bU_I^{S,M}$ with $n_I^{S,M}$ degrees of freedom and interface displacements $\bU_C^{S,M}$ with $n_C^{S,M}$ degrees of freedom.
The displacements of the master side are approximated by the product 

\begin{equation}\label{eq:masterApprox}
    \bU^M = \begin{bmatrix} \bU_I^M \\ \bU_C^M \end{bmatrix} \approx \begin{bmatrix} \BPsi_I^M  \\  \BPsi_C^M \end{bmatrix}  \ba^M  = \BPsi^M \ba^M
\end{equation}
where $\BPsi_I^M$ and $\BPsi_C^M$ are the internal and contact parts of the master-side projection matrix $\BPsi^M$ with the dimensions $n^{M}\times m^M$.
The vector $\ba^M$ contains the $m^M$ reduced displacements of the master substructure.
The product 
\begin{equation}\label{eq:slaveApprox}
    \bU^S = \begin{bmatrix} \bU_I^S \\ \bU_C^S \end{bmatrix} \approx \begin{bmatrix} \BPsi_I^S & \bm 0 \\ \bm 0 & \bI \end{bmatrix} \begin{bmatrix} \ba_I^S \\ \bU_C^S \end{bmatrix}
\end{equation}
approximates the slave-side displacements. 
Here, only the internal DOFs are approximated by the $n_I^{S}\times m_I^S$-dimensional projection matrix $\BPsi_I^S$ and the corresponding unknowns $\ba_I^S$. 
The slave-side interface displacements $\bU_C^S$ are not approximated since they can be expressed in terms of the interface-coupling operator $\bP$ and the approximation of the master-side displacements $\bU_C^M$ (cf. \Cref{eq:interfaceCoupling})
\begin{equation}\label{eq:slaveInterfaceApprox}
    \bU_C^S = \bP \,\BPsi_C^M \ba^M  
\end{equation}
The reduced system of equations is obtained by inserting the relations \Cref{eq:masterApprox}, \Cref{eq:slaveApprox} and \Cref{eq:slaveInterfaceApprox} into the condensed equation system \Cref{eq:condensedLGS} and applying a Galerkin projection with the projection matrices $\BPsi_I^M$, $\BPsi_I^S$ and $\BPsi_C^M$. 
In the Galerkin projection we multiply the transposed projection matrices from the left side to the corresponding degrees of freedom.  
This leads to the reduced quantities: 
\begin{align}
    \label{eq:reducedStiffness}
    \hat \bK_{\rm cond} &= \begin{bmatrix}
        (\BPsi^M)^T \bK^M \BPsi^M   +(\BPsi_c^M)^T \bP^T \bK_{cc}^S \bP \,  \BPsi_c^M & (\BPsi_c^M)^T \bP^T\bK_{ci}^S \BPsi_i^S  \\ 
        (\BPsi_i^M)^T \bK_{ic}^S \bP \BPsi_c^M & (\BPsi_i^S)^T \bK_{ii}^S \BPsi_i^S
        \end{bmatrix} \\ 
    \Delta \ba_{\rm cond} &= \begin{bmatrix} 
        \Delta \ba^M \\ \Delta \ba_i^S 
    \end{bmatrix}\\ 
    \hat\bG_{\rm cond} &= \begin{bmatrix} 
        (\BPsi^M)^T \bG^M + (\BPsi_c^M)^T \bP^T \bG_c^S \\ (\BPsi_i^S)^T \bG_i^S 
    \end{bmatrix}
    \label{eq:reducedResidual}
\end{align} 
In \Cref{eq:reducedStiffness} the master-side contact and internal contributions are summed up:
\begin{align} 
    \begin{aligned}
        (\BPsi^M)^T \bK^M \BPsi^M  = (\BPsi_i^M)^T \bK_{ii}^M \BPsi_i^M  + (\BPsi_i^M)^T \bK_{ic}^M \BPsi_c^M \\+ (\BPsi_c^M)^T \bK_{ci}^M \BPsi_i^M +  (\BPsi_c^M)^T \bK_{cc}^M  \BPsi_c^M 
    \end{aligned}\\ 
    (\BPsi^M)^T \bG^M = (\BPsi_i^M)^T \bG_i^M + (\BPsi_c^M)^T \bG_c^M
\end{align} 
The Lagrange Multipliers can be computed from the reduced displacements by 
\begin{equation}
    \BLambda= \bD^{-T} \left( -\bG_C^S - \bK_{CI}^S \, \BPsi_I^S \, \Delta \ba_I^S - \bK_{CC}^S \bP\, \BPsi_C^M \Delta \ba^M \right).
\end{equation}
The discrete nonlinear system of reduced equations is solved by the Newton-Raphson method.
The iterative solution algorithm for a time step is shown in \Cref{alg:NewtonRaphson}
\begin{algorithm}[hptb]
    \caption{Newton-Raphson solution of the nonlinear modular system}
    \label{alg:NewtonRaphson} 
    \SetAlgoLined
    \While{$\lVert \hat \bG_{\rm cond}(\bU^M_j,\bU_j^S) \rVert > tol.$}{
        $\Delta \ba = - \left(\hat \bK_{\rm cond}(\bU^M_j,\bU_j^S)\right)^{-1} \hat \bG_{\rm cond}(\bU^M_j,\bU_j^S) $\\ 
        $\bU^M_{j+1} = \bU^M_{j} + 
        \begin{bmatrix}
        \BPsi_I^M \Delta \ba^M \\ \BPsi_C^M \Delta \ba^M 
        \end{bmatrix} $\\
        $\bU^S_{j+1} = \bU^S_{j} + 
        \begin{bmatrix}
        \BPsi_I^S \Delta \ba_I^S \\ \bP \BPsi_C^M \Delta \ba^M 
        \end{bmatrix} $\\ 
        $j \leftarrow j+1$ \\
    }
\end{algorithm}

\paragraph{Comment}It should be noted, that for every iteration the residuals $\bG_{S}, \bG_{M}$ and the tangential stiffness matrices $\bK_{S}, \bK_{M}$ have to be computed for every substructure.  
The computational effort to solve the system is reduced but the computation of those matrices still depends on the original problem size $n = n_M+n_S$. 
At this stage hyperreduction techniques could be used to reduce this effort even further. 
In nonlinear solid mechanics the discrete empirical interpolation method (DEIM) (see  \cite{chaturantabut_nonlinear_2010,radermacher_pod-based_2016} ), (continuous) empirical cubature (\cite{hernandez_dimensional_2017,hernandez_cecm_2024}), or energy conserving sampling and weighting (ECSW) (\cite{farhat_structure-preserving_2015,rutzmoser_model_2018}) proved to be well suited.  
However, having a working POD-based model order reduction is a necessary step towards hyperreduced component-based model order reduction. 
The extension to hyperreduction is out of the scope of this paper but will be addressed in future works.

\subsubsection*{Proper orthogonal decomposition}
In the previous paragraph, we introduced the projection matrices $\BPsi_I^M$, $\BPsi_C^M$, and $\BPsi_I^S$ but have not mentioned yet how they can be computed. 
We use the proper orthogonal decomposition (POD) method to compute the projection matrices since it showed good results for nonlinear mechanics simulations (\cite{radermacher_proper_2013,rutzmoser_model_2018}). 
In POD, the projection matrix is computed from collected data of $l$ displacement states of the substructure, the so-called snapshots. 
The snapshots are stored in a matrix 
\begin{equation}
    \bm{S}_{\rm Snap} = \left[ \bm{U}_1, \bm{U}_2  , \dots, \bm{U}_l \right] =  \bm\Phi \bm\Theta \bm\Omega \qquad \: \text{(SVD)}
\end{equation}
which can be decomposed by a singular value decomposition (SVD) into the left and right mode matrix $\BPhi$ and $\BOmega$, respectively, as well as the singular values $\BTheta$.
The projection matrix is constructed by selecting only the first $m$ columns of $\BPhi$. 
Leading to the $n^s\times m$ dimensional projection matrix
\begin{equation}
    \label{eq:Psi} 
    \bm{\Psi} = \left[ \BPhi_1,  \dots , \BPhi_{m} \right],
\end{equation} 
where $n^s$ is the number of DOFs of the substructure. 
The projection matrices for the internal DOFs $\BPsi_I^M$ and $\BPsi_I^S$ or interface DOFs $\BPsi_C^M$ are obtained by selecting the corresponding DOFs of the projection matrix of a substructure.

\subsection{Computation of snapshots}\label{sec:snapshots} 

For the component-wise model order reduction technique described in \Cref{sec:ROM} suitable substructure modes are needed. 
These modes should give accurate results when the substructure is used in arbitrary structures and subjected to varying loading or tied-contact conditions. 
To compute such modes the snapshots should contain many deformation states that could possibly occur, such that the modes can be used in different structures under different loading conditions.
To achieve this, we parametrize the boundary conditions applied on the possible interface surfaces of the substructures.
The snapshots are computed by evaluating the full-order model of the substructure for different sample points in the parameter space. 
In this work, the sampling points are selected by the Latin hypercube sampling method (LHS), which creates sample points over the whole parameter space and avoids clustering. 
Other possible methods are greedy search methods, where the sample parameters are selected adaptively (see e.g. \cite{bui-thanh_model_2008} or \cite{haasdonk_reduced_2008}).  
An overview of parametric model order reduction is given in \cite{benner_survey_2015}. 

\section{Numerical Examples}\label{sec:examples}

\subsection{Example 1}\label{sec:kreuz} 

\subsubsection{Boundary value problem} 
In this numerical example, we use the same substructure in two different systems.
For both systems, we use the same mode matrices, that we computed from the snapshots generated by the Latin hypercube sampling (cf. \Cref{sec:snapshots}).
This example demonstrates that the substructure modes can be used to predict solutions of systems under various loading conditions.

\begin{figure}[hptb]
    \centering
    \includegraphics[width=0.8\textwidth]{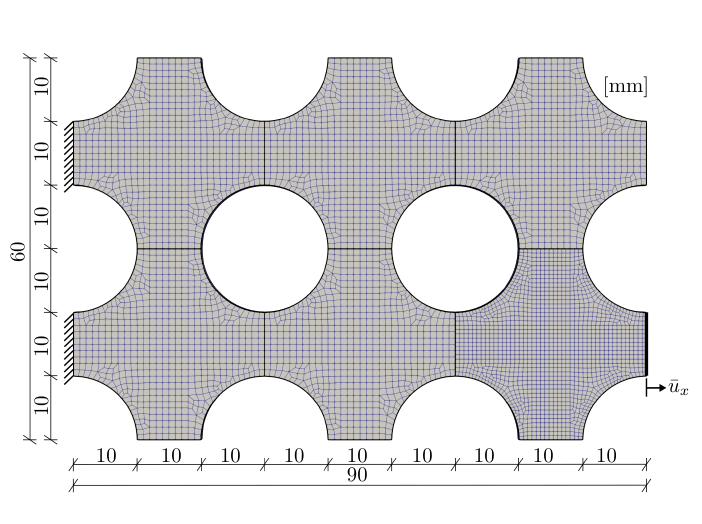}
    \caption{Geometry, mesh and boundary conditions of the $2 \times 3$ example. The substructure on the bottom right has a finer mesh than the other substructures.} 
    \label{fig:geo_2x3}
\end{figure}

The first system consists of 6 substructures and is displayed in \Cref{fig:geo_2x3}. 
The substructure in the bottom right corner has a finer mesh than the other substructures. 
In total, the system has 29358 DOFs. 
Each substructure is characterized by a Neo-Hookean material behavior with a Young's modulus of $E=80 \: \rm GPa$ and a Poisson's ratio of $\nu = 0.15$. 
In addition to this nonlinear material behavior, we also consider geometric nonlinearities by employing finite strain theory.    
Boundary conditions and dimensions are provided in \Cref{fig:geo_2x3}. 
The maximum displacement applied to the system is $\bar{u}_x^{\rm max} = 30 \: \rm mm$.

\begin{figure}[hptb]
    \centering
    \includegraphics[width=1\textwidth]{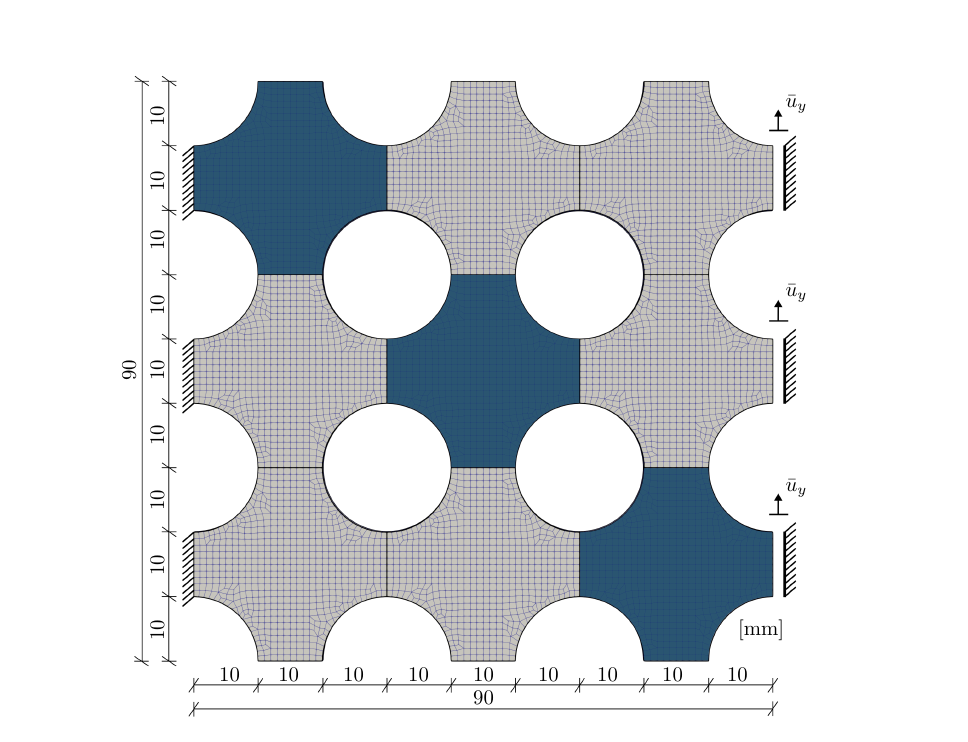}
    \caption{Geometry, mesh and boundary conditions of the $3 \times 3$ example. The blue substructures have a Young's modulus of $E=80 \: \rm GPa$, the gray substructures of $E=20 \: \rm GPa$.  } 
    \label{fig:geo_3x3}
\end{figure}

The second system consists of 9 substructures and is displayed in \Cref{fig:geo_3x3}.
Here, all substructures have the same mesh, but the substructures marked in blue have a higher stiffness than the gray substructures. 
The blue substructures have a Young's modulus of $E=80 \: \rm GPa$, the gray substructures of $E=20 \: \rm GPa$, and all substructures have the same Poisson's ratio $\nu = 0.15$. 
The system has in total 36990 DOFs. 
The boundary conditions and dimensions can be taken from \Cref{fig:geo_3x3}.   
We apply a displacement in $y$-direction on the whole right surface, while fixing the $x$-displacement on that side.

\subsubsection{Snapshot computation}\label{sec:snapsKreuz}
Snapshots are computed at the substructure level using a Latin hypercube sampling procedure, as described in \Cref{sec:snapshots}. 
The following parametrization is a specific choice, that aims to provide general snapshots that can be used to reduce arbitrary systems under varying loading conditions.
Alternative parametrizations are possible and may lead to improved results.

\begin{figure}[hptb]
    \centering
    \includegraphics[width=0.7\textwidth]{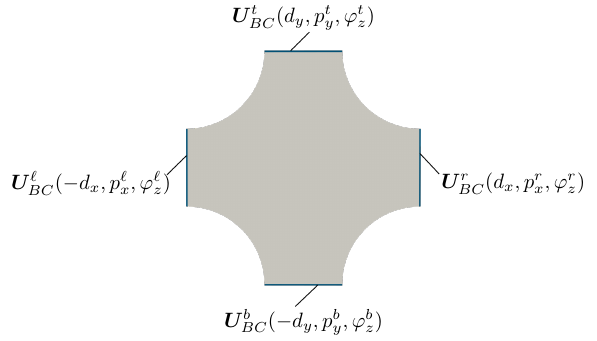}
    \caption{Illustration of the snapshot parametrization.} 
    \label{fig:parametrisierung}
\end{figure}

We parametrize the Dirichlet boundary conditions on the four edges with in total ten parameters. 
Two parameters $d_x, \, d_y$ for the normal displacements on the four edges.
The $x$-displacement on the left and right edge depend on the parameter $d_x$.
The $y$-displacement on the top and bottom edge is parametrized by $d_y$.   
Additionally, we prescribe perpendicular displacement and rotations of all four edges. 
The perpendicular displacements are described by the parameters $p_y^l$ and $p_y^r$ for the $y$-displacement on the left and right edge and $p_x^t$ and $p_y^b$ for the $x$-displacement at the top and bottom. 
The rotations of the edges lead to $x$- and $y$-displacements and are parametrized by the angles $\phi_z^l,\,\phi_z^r,\, \phi_z^t,\,\phi_z^b $. 
A snapshot is computed by applying a displacement vector at specific parameter point to a displacement driven nonlinear full-order simulation of the substructure.  
The snapshot parametrization is illustrated in \Cref{fig:parametrisierung}. 

Prescribing these displacements to the $x$- and $y$-displacements of all four edges would lead to planar surfaces without strains. 
To circumvent this problem, we compute three different snapshots with different boundary conditions for every sample point. 
In the first case, we prescribe $x$- and $y$-displacements at all DOFs of all edges. 
In the second case, we prescribe only the normal displacements of each boundary, neglecting the displacement BCs stemming from the perpendicular parameters. 
Every node at the edges can freely deform in perpendicular direction. 
In the third case we prescribe only the perpendicular displacements and let the normal displacements remain unrestricted.

The parameter space used in the LHS is defined by: 
\begin{equation*} 
    \begin{aligned} 
        &d_x \in (-1,3) \: \rm [mm] \qquad d_y\in (-1,3) \:\rm [^\circ]  \\
        &p_x^l \in (-3,3) \: \rm [mm] \qquad \phi_z^l \in (-15,15) \:\rm [^\circ]  \\ 
        &p_x^r \in (-3,3) \: \rm [mm] \qquad \phi_z^r \in (-15,15) \:\rm [^\circ]  \\ 
        &p_y^t \in (-3,3) \: \rm [mm] \qquad \phi_z^t \in (-15,15) \:\rm [^\circ]  \\
        &p_y^b \in (-3,3) \: \rm [mm] \qquad \phi_z^b \in (-15,15) \:\rm [^\circ]  
    \end{aligned}
\end{equation*} 
In this 10 dimensional parameter space we compute 50 sample points by LHS. 
At every sample points we compute two load steps for the three BC-cases described above, leading in the end to a total of 300 snapshots.   

The POD modes are computed by the singular value decomposition of the snapshot matrix consisting of the snapshots from the LHS and three snapshots containing the rigid body translations and rotations. 
The decay of the normalized singular values is shown in \Cref{fig:singular_values_kreuz} on the left side.
To assess the quality of the POD basis, we compute ten unseen test samples with increasing number of modes and analyze the mean displacement error $e$. 
The displacement error $e$ is given by 
\begin{equation}\label{eq:error}
    e = \frac{\|  \bU_{\text{MOR}} - \bU_{\text{Ref}} \| }{ \|\bU_{\text{Ref}} \|}, 
\end{equation}
where $\lVert \bullet \rVert$ denotes the $L_2$ norm of $ \bullet$.
The results can be seen on the right side of \Cref{fig:singular_values_kreuz}.
We show the mean values of the three boundary condition sets described above and the mean of all simulations. 
The blue area is the area between the minimum and maximum error.
It can be seen, that the mean displacement error drops from $e = 21.6\%$ at 14 modes to $e=3.7\%$ for 15 modes.
Utilizing more modes does not reduce the error significantly.
The mean error converges to a value of $e = 1.3\%$. 
The singular values show a similar behavior, where the slope of the decay decreases after 15 modes. 
\begin{figure}[htbp] 
    \centering
    \input{plots/singularValues_uerr_kreuz_testLHS.tex}
    \caption{Decay of the normalized singular values (left), and decay of the mean error of ten unseen test samples over the number of modes $m$. The blue area is the difference between the minimum and maximum mean displacement error.}
    \label{fig:singular_values_kreuz}
\end{figure}

The test samples show, that the POD basis can approximate the displacement states resulting from the sampling parameter space.
In the proposed method the module is used as a substructure of arbitrary systems made of multiple substructures. 
Since the substructure snapshots are not computed with those systems, the substructure POD basis is non-optimal for some displacement states that could occur in arbitrary systems. 
These systems can still be approximated if the parameter space mimics possible deformation states of the substructure. 
A higher number of modes can still increase the accuracy.      


\subsubsection{Simulation}

\begin{figure}[htbp]
    \centering
    \begin{tikzpicture}

  \definecolor{crimson2143940}{RGB}{214,39,40}
  \definecolor{darkgray176}{RGB}{176,176,176}
  \definecolor{darkorange25512714}{RGB}{255,127,14}
  \definecolor{forestgreen4416044}{RGB}{44,160,44}
  \definecolor{lightgray204}{RGB}{204,204,204}
  \definecolor{steelblue31119180}{RGB}{31,119,180}
  
  \begin{groupplot}[group style = {group name = group, group size = 2 by 1, horizontal sep = 2.2 cm}, width = 6.8 cm]  
      \nextgroupplot[
          legend cell align={left},
          legend style={font=\scriptsize,
            at={(0.03,0.97)},
            anchor=north west
          },
          title={2x3},
          xlabel={$\bar u_x\: [\rm mm]$  },
          xmin=-1.5, xmax=31.5,
          ylabel={Error $e\:[-]$},
          ymin=0.00268389440892654, ymax=0.0122565534909796,
          xmajorgrids = true, 
          ymajorgrids = true, 
          grid style = dashed
          ]
          \addplot [very thick, steelblue31119180]
          table {%
          3.75 0.00871913219846756
          7.5 0.0098128934274572
          11.25 0.0104290407730858
          15 0.0108146489515457
          18.75 0.011119778750904
          22.5 0.0113919974583649
          26.25 0.0116362548471893
          30 0.0118485208570342
          };
          \addlegendentry{$6\cdot40$}
          \addplot [very thick, darkorange25512714]
          table {%
          3.75 0.00756036432226675
          7.5 0.00799773677861938
          11.25 0.00837697571446549
          15 0.00857883391560147
          18.75 0.00864783939633617
          22.5 0.00863665828828621
          26.25 0.00858122092743305
          30 0.00850342773256632
          };
          \addlegendentry{$6\cdot50$}
          \addplot [ very thick, forestgreen4416044]
          table {%
          3.75 0.00368786817812548
          7.5 0.00414255961468195
          11.25 0.0044544660824081
          15 0.00460749081650745
          18.75 0.0046648602480848
          22.5 0.00467190303737801
          26.25 0.00465441552167666
          30 0.00462657045229385
          };
          \addlegendentry{$6\cdot60$}

      \nextgroupplot[
          legend cell align={left},
          legend style={font=\scriptsize,
            at={(0.03,0.97)},
            anchor=north west
          },
          title={3x3},
          xlabel={$\bar u_y\: [\rm mm]$  },
          xmin=-1.5, xmax=31.5,
          ylabel={Error $e \: [-]$},
          ymin=0.00268389440892654, ymax=0.0122565534909796,
          xmajorgrids = true, 
          ymajorgrids = true, 
          grid style = dashed
          ]
          \addplot [very thick, steelblue31119180]
          table {%
          3.75 0.00650018657549692
          7.5 0.00703074309486116
          11.25 0.00754057851864087
          15 0.0079549000338175
          18.75 0.00826565512068318
          22.5 0.008502336450035
          26.25 0.008699678001057
          30 0.0088809748268036
          };
          \addlegendentry{$9\cdot40$ }
          \addplot [very thick, darkorange25512714]
          table {%
          3.75 0.00571274942777267
          7.5 0.00520244055067347
          11.25 0.00473614878906158
          15 0.00431665281810028
          18.75 0.00395982591399492
          22.5 0.00369307888072675
          26.25 0.00353895982225877
          30 0.00350307089535811
          };
          \addlegendentry{$9\cdot50$ }
          \addplot [very thick, forestgreen4416044]
          table {%
          3.75 0.00619990657718355
          7.5 0.00568454701584271
          11.25 0.0050021317775983
          15 0.00430563393698713
          18.75 0.00371251135001125
          22.5 0.00328601647270925
          26.25 0.00304574874997578
          30 0.0029789934764445
          };
          \addlegendentry{$9\cdot60$ }
  \end{groupplot}

\end{tikzpicture}
    \caption{Plot of the displacement error norm over the prescribed displacement for the $2\times 3$ and $3 \times 3 $ example, for different numbers of substructure modes. For both systems, we show solutions with 40, 50, and 60 modes per substructure. }
    \label{fig:kreuz_twoplots_uerr}
\end{figure}

The accuracy of the substructuring method and substructure modes is evaluated by comparing reduced solutions with full-order mortar tied-contact simulations.
In \Cref{fig:kreuz_twoplots_uerr} displacement error norms are shown over the applied displacement for different numbers of modes per substructure. 
In general, the error of the $2\times 3$ structure is higher than that of the $3\times 3$ structure.
For all shown number of substructure modes the error is below $1.2 \%$. 
It can be seen, that by increasing the number of substructure modes the displacement error decreases. 
However, for the $3\times3$ system the displacement error of the simulation with 60 modes per substructure is not in every solution step smaller than the simulation with 50 modes per substructure. 
For the $2\times3$ system the error decreases when the number of modes is increased from 50 to 60. 

In \Cref{fig:kreuz_twoplots} we show normalized force-displacement diagrams of the two load cases, for an increasing number of substructure modes. 

For the system consisting of $2\times3$ substructures it can be seen that with all three different numbers of modes per substructure, the reference solution can be predicted. 
For the system with $3\times3$ substructures, the reaction forces of the reduced solution are overestimated. 
By increasing the number of modes of the substructure the reaction forces converge to the reference solution. 
A reason for the overestimation could be that the snapshots are computed with a Young's modulus of $E=80 \: \rm{GPa}$ and the softer substructures have a Young's modulus of $E=20 \: \rm{GPa}$.

\begin{figure}[htbp]
    \centering
    \begin{tikzpicture}

  \definecolor{crimson2143940}{RGB}{214,39,40}
  \definecolor{darkgray176}{RGB}{176,176,176}
  \definecolor{darkorange25512714}{RGB}{255,127,14}
  \definecolor{forestgreen4416044}{RGB}{44,160,44}
  \definecolor{lightgray204}{RGB}{204,204,204}
  \definecolor{steelblue31119180}{RGB}{31,119,180}
  
  \begin{groupplot}[group style = {group name = group, group size = 2 by 1, horizontal sep = 2.2 cm}, width = 6.8 cm]  
      \nextgroupplot[
          legend cell align={left},
          legend style={font=\scriptsize,
            at={(0.03,0.97)},
            anchor=north west
          },
          title={2x3},
          xlabel={$\bar u_z\: [\rm mm]$  },
          xmin=-1.5, xmax=31.5,
          ylabel={\(\displaystyle F/F_{max} \) [-]},
          ymin=-0.0506593513976169, ymax=1.09,
          xmajorgrids = true, 
          ymajorgrids = true, 
          grid style = dashed
          ]
          \addplot [very thick, steelblue31119180]
          table {%
          0 0
          3.75 0.118324563698409
          7.5 0.244231922837387
          11.25 0.372298098279716
          15 0.500377029587037
          18.75 0.62753626658942
          22.5 0.753326006826957
          26.25 0.877517748413809
          30 1
          };
          \addlegendentry{Reference}
          \addplot [very thick, dashed, darkorange25512714]
          table {%
          0 0
          3.75 0.118837675590777
          7.5 0.245356715955328
          11.25 0.374181408431092
          15 0.503149348551059
          18.75 0.631298157988804
          22.5 0.758148724276778
          26.25 0.883448794404666
          30 1.00706952306468
          };
          \addlegendentry{$6\cdot 40$}
          \addplot [very thick, dashed, forestgreen4416044]
          table {%
          0 0
          3.75 0.118638082644891
          7.5 0.244889146349175
          11.25 0.373364629963694
          15 0.501917497675411
          18.75 0.629607822800583
          22.5 0.755977011763691
          26.25 0.880788029098819
          30 1.003921732088
          };
          \addlegendentry{$6\cdot 50$}
          \addplot [very thick, dashed, crimson2143940]
          table {%
          0 0
          3.75 0.118579460077597
          7.5 0.244782797078464
          11.25 0.373197526536653
          15 0.501668406773917
          18.75 0.629252511257197
          22.5 0.755490815903257
          26.25 0.880147283590695
          30 1.00310454689693
          };
          \addlegendentry{$6\cdot 60$}

      \nextgroupplot[
          legend cell align={left},
          legend style={font=\scriptsize,
            at={(0.03,0.97)},
            anchor=north west
          },
          title={3x3},
          xlabel={$\bar u_y\: [\rm mm]$  },
          xmin=-1.5, xmax=31.5,
          ylabel={\(\displaystyle F/F_{max} \) [-]},
          ymin=-0.0506593513976169, ymax=1.1,
          xmajorgrids = true, 
          ymajorgrids = true, 
          grid style = dashed
          ]
          \addplot [very thick, steelblue31119180]
          table {%
          0 0
          3.75 0.118410579261156
          7.5 0.234661732112273
          11.25 0.350971138434636
          15 0.469461375767693
          18.75 0.592072148756673
          22.5 0.720490794400862
          26.25 0.856107132132669
          30 1
          };
          \addlegendentry{Reference}
          \addplot [very thick,dashed, darkorange25512714]
          table {%
          0 0
          3.75 0.120283146485129
          7.5 0.238819866734495
          11.25 0.358202801989494
          15 0.480848861724341
          18.75 0.608873034128082
          22.5 0.744013138204979
          26.25 0.887607770069398
          30 1.0406156400533
          };
          \addlegendentry{$9\cdot40$}
          \addplot [very thick,dashed, forestgreen4416044]
          table {%
          0 0
          3.75 0.119346385287647
          7.5 0.236637831371311
          11.25 0.354235180578791
          15 0.474395099718107
          18.75 0.599166156894367
          22.5 0.730308401532786
          26.25 0.869246961505445
          30 1.01706300292364
          };
          \addlegendentry{$9\cdot50$}
          \addplot [very thick,dashed, crimson2143940]
          table {%
          0 0
          3.75 0.11916484029856
          7.5 0.236255982853952
          11.25 0.35357391908284
          15 0.473331879415378
          18.75 0.597560199539472
          22.5 0.728024065315059
          26.25 0.866167540621041
          30 1.01309333195151
          };
          \addlegendentry{$9\cdot 60$}
          
  \end{groupplot}

\end{tikzpicture}
    \caption{Plot of the reaction forces over the displacements for the $2\times 3$ and $3 \times 3 $ example. For both systems, we show solutions with 40, 50, and 60 modes per substructure.}
    \label{fig:kreuz_twoplots}
\end{figure}

\begin{figure}[hptb]
    \centering
    \includegraphics[width=0.7\textwidth]{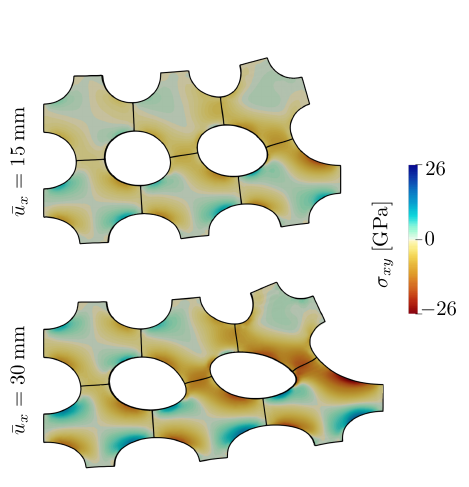}
    \caption{Cauchy shear stress contour plots of the deformed reduced solution. The black outlines are the reference solutions.} 
    \label{fig:stressXY_3x2}
\end{figure}  

\begin{figure}[hptb]
    \centering
    \includegraphics[width=1.\textwidth]{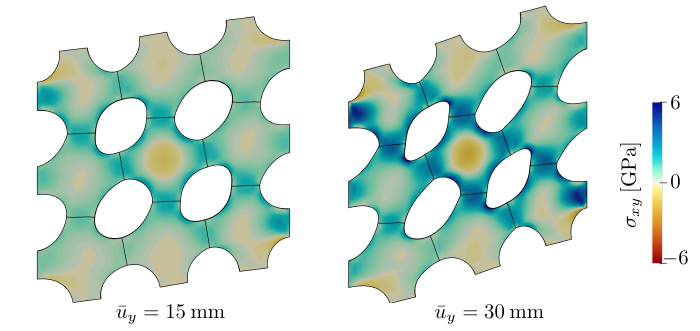}
    \caption{Cauchy shear stress contour plots of the deformed reduced solution. The black outlines are the reference solutions.} 
    \label{fig:stressXY_3x3}
\end{figure}

In \Cref{fig:stressXY_3x2} and \Cref{fig:stressXY_3x3} we show the shear stress contour plots as well as the displacements of the two systems. 
The black outline is the outline of the reference solution computed by the mortar tied-contact method without model order reduction. 
For the reduced solution of both systems 50 modes per substructure are used.
The different displacement states of the substructures in the two systems can all be described by the same substructure modes.
The modes can be used to reduce substructures with different numbers of free edges, contact interfaces or interfaces, where boundary conditions are applied.
This indicates the versatility of the computed modes.

The reduced order system has approximately 98 times fewer DOFs than the reference model $\left(\dfrac{29358}{300} \approx 98 \right)$.
The results from \Cref{fig:stressXY_3x2} show further, that non-matching meshes are possible to solve with the proposed ROM. 
The system has approximately 82 times fewer DOFs than the reference solution $\left(\dfrac{36990}{690} \approx 82 \right)$.
It can also be seen, that the modes computed with the method described in \Cref{sec:snapshots} can be used for different stiffnesses of the substructures.

\begin{figure}[hptb]
    \centering
    \includegraphics[width=1.\textwidth]{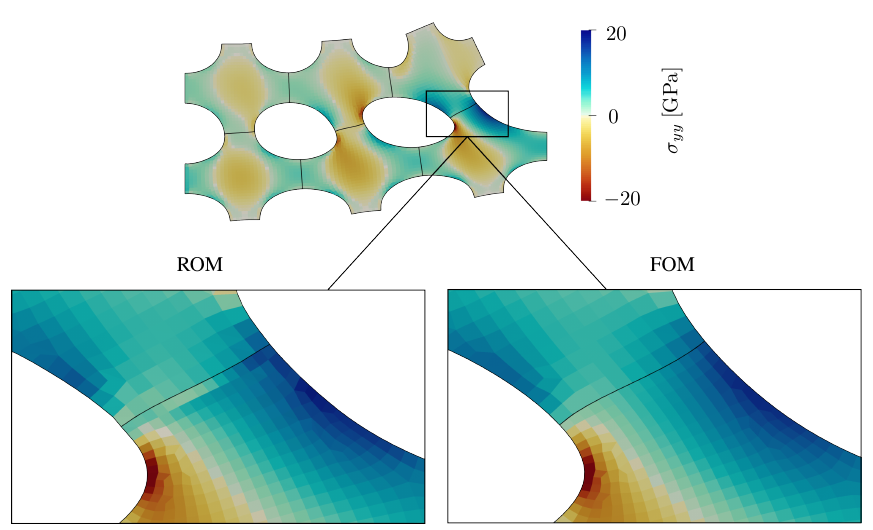}
    \caption{Zoom of the Cauchy stress $\sigma_{yy}$ contour plots at the interface for the reduced solution (left) and full-order solution (right). The black outlines are the substructure boundaries. The stresses are computed in the center of the element.  }
    \label{fig:stressYY_3x2_zoom}
\end{figure}

In \Cref{fig:stressYY_3x2_zoom} we show the $yy$-component of the Cauchy stress tensor $\Bsigma$ at the displayed interface for the ROM and FOM. 
The black outline indicates the substructure interface. 
It can be seen that the interface curvature of the ROM is smaller than of the FOM. 
Consequently, the stresses in the interface region differ. 
With increasing distance from the interface the stresses are approximated more accurate.

\subsection{Example 2: Ring-segment}\label{sec:schlange} 
\subsubsection{Boundary value problem} 
\begin{figure}[ht]
    \centering
    \def\svgwidth{0.8\textwidth}
    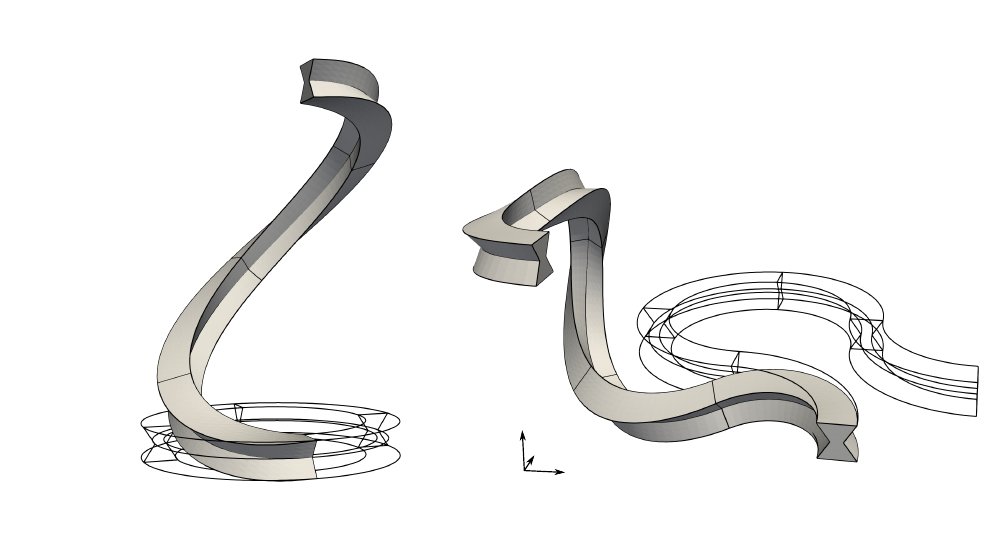
    \caption{Boundary value problems of two systems "ring" and "omega", that are assembled from the same modules. For both systems, the undeformed configuration is shown as an outline, and the deformed configuration is shown in gray. Both systems are fixed where the deformed and the undeformed configurations coincide. The displacement boundary conditions are applied to the marked area.  }
    \label{fig:geometrySchlange}
\end{figure}

\begin{figure}[ht]
    \centering
    \def\svgwidth{0.7\textwidth}
    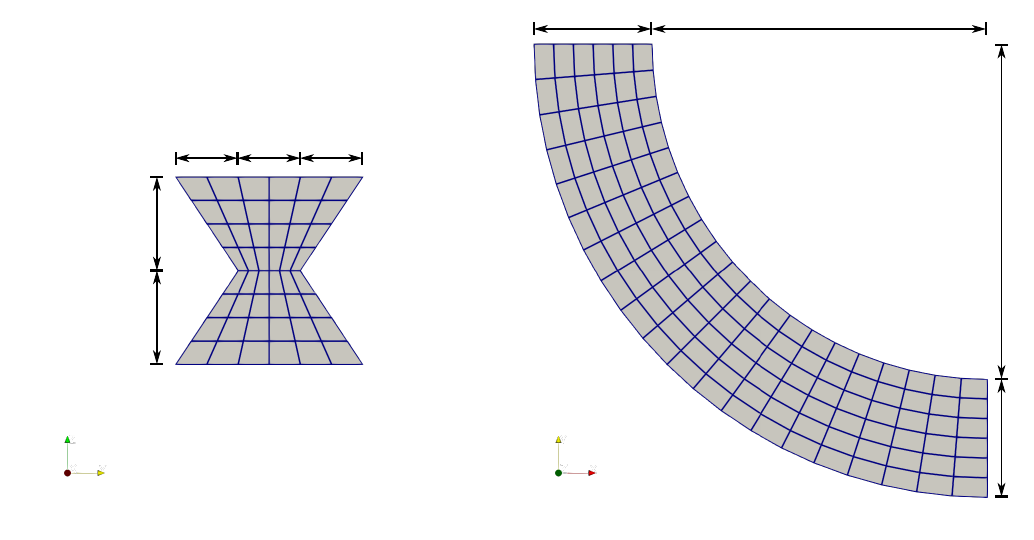
    \caption{Geometry and mesh of a module. Cross-section (left) and top view (right)}
    \label{fig:meshSchlange}
\end{figure}

In this numerical example, we use the same substructure in two different systems.
The dimensions and mesh of the component are shown in \Cref{fig:meshSchlange}.
The boundary value problems are illustrated in \Cref{fig:geometrySchlange}. 
On the left, four components are assembled into a ring structure, with 15876 DOFs in total.  
The other system consists of five substructures with 19845 DOFs. 
The displacement boundary conditions can be seen in \Cref{fig:geometrySchlange}.
In the following, we use the term "ring"-structure for the example on the left and the term "omega" for the structure shown on the right.   

In \Cref{sec:Schlange_neoHooke} the results for a Neo-Hookean material and in \Cref{sec:Schlange_visco} the results for a finite strain viscoelasticity model are shown.
For all simulations we use mode matrices computed from the same set of snapshots. 
The snapshot parametrization is explained in \Cref{sec:ring_snaps}.

In \Cref{fig:meshSchlange} we show the meshes of the module used in the numerical examples from \Cref{fig:geometrySchlange}. 

\subsubsection{Snapshot computation}\label{sec:ring_snaps} 
\begin{figure}[hptb]
    \centering
    \includegraphics[width=0.85\textwidth]{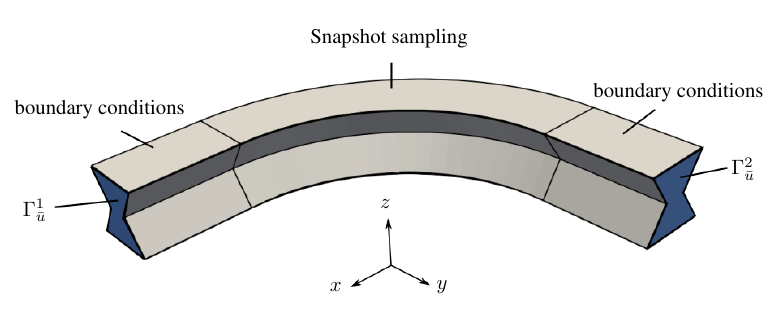}
    \caption{Boundary value problem for the snapshot computation, consisting of three substructures. The snapshots are collected for the central substructure.} 
    \label{fig:snapshot_ringstueck}
\end{figure}

The snapshots are computed for the substructure in the middle with the boundary value problem shown in \Cref{fig:snapshot_ringstueck}. 
The two substructures, that are attached left and right are used to apply the boundary conditions on the surfaces $\Gamma_{\bar u}^1$ and $\Gamma_{\bar u}^2$. 
This makes deflections of the interfaces in $x$- and $y$-direction possible. 
These deflections are necessary, because these interfaces do not remain planar after deformation if the substructures are assembled into larger systems.

On both surfaces, we apply a displacement that depends on six parameters \(\bU_{BC} = \bU_{BC}(d_x, \, d_y, \, \\ d_z,  \,\phi_x, \, \phi_y, \, \phi_z ) \). 
The parameters \( d_x, \, d_y, \, d_z \) describe a displacement of the surface in $x,y,z$-direction. 
The parameters \( \phi_x, \, \phi_y, \, \phi_z \) rotate the surfaces $\Gamma_{\bar u }^1$ and $\Gamma_{\bar u }^1$ around the $x,y,z$-axes. 
The parameter ranges for the Latin hypercube sampling (LHS) are:
\begin{equation*} 
    \begin{aligned} 
        d_x \in (-80,80) \: \rm [mm] \qquad \phi_x \in (-60,60) \:\rm [^\circ]  \\
        d_y \in (-30,30) \:\rm [mm] \qquad \phi_y \in (-60,60) \: \rm [^\circ]  \\
        d_z \in (-100,100)\: \rm [mm] \qquad \phi_z \in (-15,15) \: \rm [^\circ]  \\
    \end{aligned}
\end{equation*}
For each random parameter point, we compute two snapshots. 
One where we apply the displacement to the surface $\Gamma_{\bar u }^1$ and set all displacement of the surface $\Gamma_{\bar u }^2$ to zero, and a second one where the same displacement is applied to $\Gamma_{\bar u }^2$.
For the sampling we use a Neo-Hookean material model with the Lame constants: $\lambda = 14907 \:\rm MPa, \mu = 34783 \: \rm MPa$. 
We compute snapshots at 60 LHS sample points where the displacement boundary condition is applied incrementally in three load steps. 
Three load steps are necessary to ensure convergence for all parameter combinations. 
Additionally, we compute snapshots with the minimal and maximal value of each parameter, while all other parameters are zero. 
This ensures, that the extreme values of the parameter space are covered in the snapshots. 
Leading in total to a number of 486 snapshots, including also the six rigid body motions.

The POD basis is computed by the singular value decomposition. 
The decay of the normalized singular values can be seen in \Cref{fig:singular_values_ring}. 
Additionally, we generate 10 test samples by Latin hypercube sampling that where not included in the snapshot samples. 
We analyzed the mean displacement error of this test samples for reduced simulations with increasing number of modes. 
In the simulations only the substructure in the center is reduced, the pieces at the boundary are unreduced and are used to compute apply the boundary conditions to the center substructure. 

\begin{figure}[htbp] 
    \centering
    \input{plots/singularValues_uerr_ring_testLHS.tex}
    \caption{Decay of the normalized singular values (left), and decay of the mean error of ten unseen test samples over the number of modes $m$. The blue area is the difference between the minimum and maximum mean displacement error.}
    \label{fig:singular_values_ring}
\end{figure}

The mean displacement error $e$ decreases with increasing number of modes. 
With 40 modes for the ring substructure the maximum error is $1.1\%$. 
More modes lead to a further decrease of the error. 
For 100 modes per substructure the maximum error is $0.2\%$. 
In contrast to the displacement error shown in \Cref{fig:singular_values_kreuz} (\Cref{sec:snapsKreuz}) it does not converge. 
A difference between those two systems is, that the reduced test sample simulations of the "ring" substructure are computed with the method described in \Cref{sec:ROM}. 
With the modification, that only the substructure in the center was reduced. 
The other two substructures where unreduced, because only the POD approximation of the substructure in the center should be tested.
It could be possible that the different error behavior is related to this difference.

\subsubsection{Neo-Hooke material}\label{sec:Schlange_neoHooke} 
First, we investigate the accuracy of the substructuring MOR method and the modes for a Neo-Hookean material. 
The modes are computed from the snapshots described above in \Cref{sec:ring_snaps}. 
In \Cref{fig:contour_ring,fig:contour_omega} in \Cref{apx:ring_omega} we show the deformed structures and the stress contours computed by the here proposed method.

In \Cref{fig:ring_uerr} we compare the accuracy of the displacement field for an increasing number of modes. 
We also compare the method to the penalty approach proposed in \cite{zhou_proper_2018}. 
It can be seen, that the proposed method (cf. \Cref{sec:ROM}) is more accurate than the solutions computed by the penalty method. 
The error norm $e$ for the proposed method is for all three number of modes below $0.15 \%$ (the displacement error is defined in \Cref{eq:error}). 
Increasing the number of modes to 80 modes per substructure can reduce the error to $0.05 \%$.

For the penalty method the most accurate results are computed with 60 modes per substructure. 
Increasing the number of modes here leads to increasing errors. 
The reason for this behavior is, that the solution depends on the choice of the penalty parameter. 
If the penalty parameter is too high, the displacements are underestimated. For too small values the reaction forces are overestimated. 
The accuracy could be increased by choosing a higher penalty parameter. 
The parameter not only influences the accuracy but also the numerical stability.
In this example, the penalty parameter was chosen as  $\epsilon = 10^3$ to get converging results. 
For this parameter, we needed 10 times more load steps to reach convergence.
For higher penalty parameters the reduced solution did not converge at all.  
The reduced solution underestimates the displacements.
If the small penalty parameter overestimates the displacement this can lead to smaller errors even though fewer modes are used.

\begin{figure}[htbp]
    \centering
    \begin{tikzpicture}

    \definecolor{crimson2143940}{RGB}{214,39,40}
    \definecolor{darkgray176}{RGB}{176,176,176}
    \definecolor{darkorange25512714}{RGB}{255,127,14}
    \definecolor{forestgreen4416044}{RGB}{44,160,44}
    \definecolor{lightgray204}{RGB}{204,204,204}
    \definecolor{steelblue31119180}{RGB}{31,119,180}
    
    \begin{groupplot}[group style = {group name = group, group size = 2 by 2, horizontal sep = 2.2 cm, vertical sep = 2.2 cm}, width = 6.8 cm]  
        \nextgroupplot[
            legend cell align={left},
            legend style={font=\scriptsize,
              at={(0.03,0.97)},
              anchor=north west
            },
            title={MOR - Mortar},
            xlabel={$\bar u_z\: [\rm mm]$  },
            xmin=-15, xmax=315,
            ylabel={Error $e  \:\: [-] $},
            ymin=-0.000223211347657164, ymax=0.0102023897073803,
            xmajorgrids = true, 
            ymajorgrids = true, 
            grid style = dashed
            ]
            \addplot [very thick, darkorange25512714]
            table {%
            37.5 0.00035121931830794
            75 0.000423997967447697
            112.5 0.000643467842619474
            150 0.00085036354729845
            187.5 0.00100739597600205
            225 0.00112700553018166
            262.5 0.00124952785065521
            300 0.00142817132327531
            };
            \addlegendentry{$4\cdot60$}
            \addplot [very thick, forestgreen4416044]
            table {%
            37.5 0.000155810713315257
            75 0.000269005949110295
            112.5 0.000521803090228436
            150 0.000779469312007746
            187.5 0.00100290679206846
            225 0.00118019280032223
            262.5 0.00132087058716873
            300 0.00145418629236495
            };
            \addlegendentry{$4\cdot70$}
            \addplot [very thick, crimson2143940]
            table {%
            37.5 0.000141656238792422
            75 0.000159210004735219
            112.5 0.00021534486546686
            150 0.000292493899104154
            187.5 0.000371595025279204
            225 0.000434091121976783
            262.5 0.000466429402650632
            300 0.000477264675651058
            };
            \addlegendentry{$4\cdot80$}
        \nextgroupplot[
            legend cell align={left},
            legend style={font=\scriptsize,
              at={(0.03,0.97)},
              anchor=north west
            },
            title={MOR - Penalty},
            xlabel={$\bar u_z\: [\rm mm]$  },
            xmin=-15, xmax=315,
            ylabel={Error $e  \:\: [-] $},
            ymin=-0.000223211347657164, ymax=0.0102023897073803,
            xmajorgrids = true, 
            ymajorgrids = true, 
            grid style = dashed
            ]
            \addplot [very thick, darkorange25512714]
            table {%
            37.5 0.00019721084020098
            75 0.000485005352019848
            112.5 0.00118856885932827
            150 0.00199795812570049
            187.5 0.00284989679021226
            225 0.00370890941073297
            262.5 0.0045497987447988
            300 0.00534043329806389
            };
            \addlegendentry{$4\cdot60$}
            \addplot [very thick, forestgreen4416044]
            table {%
            37.5 0.000556478260864238
            75 0.00133778943805786
            112.5 0.00245212394490891
            150 0.00379257161865061
            187.5 0.00523648481641469
            225 0.00667560754324764
            262.5 0.00802577537927783
            300 0.00922740557902285
            };
            \addlegendentry{$4\cdot70$}
            \addplot [very thick, crimson2143940]
            table {%
            37.5 0.000558629963193031
            75 0.00134861506185968
            112.5 0.00246386006858036
            150 0.00380884069571274
            187.5 0.00526515713457578
            225 0.00672744029579517
            262.5 0.00811304778317713
            300 0.00936237283428582
            };
            \addlegendentry{$4\cdot80$}

    \end{groupplot}

\end{tikzpicture}
    \caption{Plot of the displacement error norm over the prescribed displacement of the "ring" example. On the left the contact conditions are enforced by the reduced mortar approach and on the right by the penalty method. For both methods we show solutions with 60, 70, and 80 modes per substructure.}
    \label{fig:ring_uerr}
\end{figure}

Similar observations can also be made for the force displacement curves shown in \Cref{fig:ring_force_disp}. 
The curves computed by the mortar MOR method converge to the reference solution by increasing the number of modes. 
For the penalty method the best approximation of the force displacement curve is the simulation with 60 modes per substructure. 
Increasing the number of modes leads to a underestimation of the reaction forces. 
The reason is the influence of the penalty parameter on the solution.

\begin{figure}[htbp]
    \centering
    \input{plots/ring_force_disp_twoplots.tex}
    \caption{Force displacement curve of the "ring" example for the two different approaches that enforce the tied contact condition. The mortar approach is shown on the left and the penalty approach on the right. For both methods we show solutions with 60, 70, and 80 modes per substructure. }
    \label{fig:ring_force_disp}
\end{figure}

Analogously to \Cref{fig:ring_uerr}, \Cref{fig:omega_uerr} shows the displacement error norm over the applied total displacement $\sqrt{\bar u_x^2+ \bar u_z^2} $. 
The error norms of the mortar method and the penalty method are compared for increasing numbers of modes. 
Compared to the "ring" example more modes have to be included in order to get accurate results.   
For 80 modes per substructure the maximum error for the "omega" example is one order of magnitude higher than the error of the "ring" example. 
Increasing the number of modes can reduce this error. 
For 120 modes per substructure the maximum error is $e_{\rm max} = 0.0025$.
The penalty solution is less accurate than the mortar solution. 
For 120 modes per substructure the maximum error is here $e_{\rm max} = 0.0041$.

\begin{figure}[htbp]
    \centering
    \begin{tikzpicture}

    \definecolor{crimson2143940}{RGB}{214,39,40}
    \definecolor{darkgray176}{RGB}{176,176,176}
    \definecolor{darkorange25512714}{RGB}{255,127,14}
    \definecolor{forestgreen4416044}{RGB}{44,160,44}
    \definecolor{lightgray204}{RGB}{204,204,204}
    \definecolor{steelblue31119180}{RGB}{31,119,180}
    
    \begin{groupplot}[group style = {group name = group, group size = 2 by 2, horizontal sep = 2.2 cm, vertical sep = 2.2 cm}, width = 6.8 cm]  
        \nextgroupplot[
            legend cell align={left},
            legend style={font=\scriptsize,
              at={(0.03,0.97)},
              anchor=north west
            },
            title={MOR - Mortar},
            xlabel={$\sqrt{ \bar u_x^2 + \bar u_z^2}\:\: [\rm mm] $  },
            xmin=-16.7705098312484, xmax=352.180706456217,
            ylabel={Error $e  \:\: [-] $},
            ymin=-0.000326958273132175, ymax=0.0196654434073191,
            xmajorgrids = true, 
            ymajorgrids = true, 
            grid style = dashed
            ]
            \addplot [very thick, darkorange25512714]
            table {%
            20.9375 0.000929429615093788
            41.875 0.000937640883219619
            62.8125 0.00104749320587475
            83.75 0.00134651850925339
            104.6875 0.00188621668916926
            125.625 0.00263332829534103
            146.5625 0.00345277611476723
            167.5 0.00416979587192437
            188.4375 0.00479107241934159
            209.375 0.0055265658932732
            230.3125 0.00639583266837804
            251.25 0.00715768133438556
            272.1875 0.00759886011711761
            293.125 0.0076487736436236
            314.0625 0.00734210886349151
            335 0.0067669649224781
            };
            \addlegendentry{$5\cdot80$}
            \addplot [very thick, forestgreen4416044]
            table {%
            20.9375 0.000673832013884048
            41.875 0.000658753373785165
            62.8125 0.00065003564706678
            83.75 0.000688336587209865
            104.6875 0.000823968458727078
            125.625 0.00106616523438065
            146.5625 0.00136013564503885
            167.5 0.00167584964692903
            188.4375 0.00219083277292717
            209.375 0.0030293891289959
            230.3125 0.00390418542280497
            251.25 0.00449311009875173
            272.1875 0.00468127564187923
            293.125 0.00450218574493025
            314.0625 0.00404693558616382
            335 0.00341222693423269
            };
            \addlegendentry{$5\cdot100$}
            \addplot [very thick, crimson2143940]
            table {%
            20.9375 0.00057465743163228
            41.875 0.000558570387015083
            62.8125 0.000535644406261734
            83.75 0.000515404928336512
            104.6875 0.000519028396448646
            125.625 0.000569624858072741
            146.5625 0.000665544100605001
            167.5 0.000819448410817849
            188.4375 0.00113505941687599
            209.375 0.00162008403080555
            230.3125 0.00210099682430701
            251.25 0.00242649907353191
            272.1875 0.00254507671114904
            293.125 0.00247191635302694
            314.0625 0.0022506704138725
            335 0.00192961568053715
            };
            \addlegendentry{$5\cdot120$}
        \nextgroupplot[
            legend cell align={left},
            legend style={font=\scriptsize,
              at={(0.03,0.97)},
              anchor=north west
            },
            title={MOR - Penalty},
            xlabel={$\sqrt{ \bar u_x^2 + \bar u_z^2}\:\: [\rm mm] $  },
            xmin=-16.7705098312484, xmax=352.180706456217,
            ylabel={Error $e  \:\: [-] $},
            ymin=-0.000326958273132175, ymax=0.0196654434073191,
            xmajorgrids = true, 
            ymajorgrids = true, 
            grid style = dashed
            ]
            \addplot [very thick, darkorange25512714]
            table {%
            20.9375 0.00379753183675937
            41.875 0.00391145168758887
            62.8125 0.00506265254070292
            83.75 0.00698723562416705
            104.6875 0.00949113666307232
            125.625 0.0122177063557158
            146.5625 0.014609771026033
            167.5 0.0162462687696245
            188.4375 0.017304322342794
            209.375 0.0181725350806939
            230.3125 0.0187566978763895
            251.25 0.018725974263302
            272.1875 0.0179398035969212
            293.125 0.0164737706397141
            314.0625 0.0145047876907745
            335 0.0122304945561745
            };
            \addlegendentry{$5\cdot80$}
            \addplot [very thick, forestgreen4416044]
            table {%
            20.9375 0.000581787257797428
            41.875 0.00063988915887681
            62.8125 0.000836775798517276
            83.75 0.00131159099383129
            104.6875 0.00201549017425733
            125.625 0.00282285604335953
            146.5625 0.00355013530163923
            167.5 0.00415107213855567
            188.4375 0.00507036074091076
            209.375 0.00659515432071787
            230.3125 0.00814709082780505
            251.25 0.00912717149302346
            272.1875 0.00937475507432584
            293.125 0.00900791865555313
            314.0625 0.00823059478583734
            335 0.0072357081824327
            };
            \addlegendentry{$5 \cdot 100 $} 
            \addplot [very thick, crimson2143940]
            table {%
            20.9375 0.000621946307098946
            41.875 0.000623641410085403
            62.8125 0.000618272289863386
            83.75 0.000707124466367225
            104.6875 0.000935153151061305
            125.625 0.00124354882314897
            146.5625 0.00153484829485673
            167.5 0.00185545999754016
            188.4375 0.00248532843586449
            209.375 0.00337648170881211
            230.3125 0.00411165062487541
            251.25 0.00443494276316137
            272.1875 0.00433397727316841
            293.125 0.00391071500442143
            314.0625 0.00328819604578748
            335 0.00257224089843459
            };
            \addlegendentry{$5 \cdot 120 $}

    \end{groupplot}

\end{tikzpicture}
    \caption{ Plot of the displacement error norm over the prescribed displacement of the "omega" example. On the left the contact conditions are enforced by the reduced mortar approach and on the right by the penalty method. For both methods we show solutions with 80, 100, and 120 modes per substructure. }
    \label{fig:omega_uerr}
\end{figure}

In \Cref{fig:omega_force_disp} we show the force-displacement curves for the "omega" structure. 
From the force-displacement curve it can be seen that the system shows stability effects. 
At a displacement of around $\sqrt{\bar u_x^2+ \bar u_z^2} \approx 200 \, \rm mm $ the system can deform without increased resistance up to a displacement of $\sqrt{\bar u_x^2+ \bar u_z^2} \approx 250 \, \rm mm $, where the reaction forces increase again. 
The reduced simulation can qualitatively predict this behavior, but the reaction forces are overestimated.
The results of the mortar method are closer to the reference solution than the results of the penalty method. 
For 120 modes per substructure the force error of the mortar method is $(F_{\rm mortar} - F_{\rm Ref})/ F_{\rm Ref} = 0.035$. 
The force error of the penalty method is nearly twice as high with $(F_{\rm penalty} - F_{\rm Ref})/ F_{\rm Ref} = 0.068$.

\begin{figure}[htbp]
    \centering
    \input{plots/omega_NH_force_disp_twoplots.tex}
    \caption{Force displacement curve of the omega-shaped example for the two different approaches that enforce the tied contact condition. The mortar approach is shown on the left and the penalty approach on the right.
    For both methods we show solutions with 80, 100, and 120 modes per substructure. }
    \label{fig:omega_force_disp}
\end{figure}

In \Cref*{fig:both_times} we show the relative simulation times regarding the total simulation time of the FOM. 
For the "ring" structure the simulation time is $33.12\, \%$ of the FOM time and for the "omega" structure it is $37.47 \,  \%$. 
In \Cref{fig:both_times} we split the simulation times into two parts. 
Firstly, the assembly of the system and, secondly, the solution of the system. 
Most time savings were achieved in the solution of the system since the dimension of the ROM is smaller than the dimension of the FOM. 
The assembly of the system cannot be sped up by the current reduced model, because in every iteration each element stiffness matrix and residual has to be called and projected into the reduced space. 
Hyperreduction techniques like ECSW could reduce this computational effort. 

\begin{figure}[htbp]
    \centering
    \begin{tikzpicture}

    \definecolor{crimson2143940}{RGB}{214,39,40}
    \definecolor{darkgray176}{RGB}{176,176,176}
    \definecolor{darkorange25512714}{RGB}{255,127,14}
    \definecolor{forestgreen4416044}{RGB}{44,160,44}
    \definecolor{lightgray204}{RGB}{204,204,204}
    \definecolor{steelblue31119180}{RGB}{31,119,180}
    
    \begin{groupplot}[group style = {group name = group, group size = 2 by 1, horizontal sep = 2.2 cm}, width = 6.8 cm]  
        \nextgroupplot[grid style = dashed,
          xmajorgrids = false,
          ymajorgrids = true, 
          title={"Ring"},  
          legend cell align={left},
          legend style={font=\scriptsize},
          tick align=outside,
          tick pos=left,
          x grid style={darkgray176},
          xmajorgrids,
          xmin=-0.38, xmax=1.38,
          xtick style={color=black},
          xtick={0,1},
          xticklabels={FOM,ROM},
          y grid style={darkgray176},
          ymajorgrids,
          ymin=0, ymax=1.05,
          ytick style={color=black},
          ylabel = {relative simulation time $[-]$},
          ybar,
          legend image code/.code={\draw [#1,draw=none] (0cm,-0.1cm) rectangle (0.2cm,0.12cm); }
          ]
          \draw[draw=none,fill=steelblue31119180] (axis cs:-0.3,0) rectangle (axis cs:0.3,0.370700173330561);
          \addlegendimage{draw=none,fill=steelblue31119180}
          \addlegendentry{Assembly}

          \draw[draw=none,fill=steelblue31119180] (axis cs:0.7,0) rectangle (axis cs:1.3,0.322218618705329);
          \draw[draw=none,fill=darkorange25512714] (axis cs:-0.3,0.370700173330561) rectangle (axis cs:0.3,1);
          \addlegendimage{draw=none,fill=darkorange25512714}
          \addlegendentry{Solution}

          \draw[draw=none,fill=darkorange25512714] (axis cs:0.7,0.322218618705329) rectangle (axis cs:1.3,0.331205786802528);
        \nextgroupplot[
          grid style = dashed,
          xmajorgrids = false,
          ymajorgrids = true, 
          title={"Omega"},
          legend cell align={left},
          legend style={font=\scriptsize},
          tick align=outside,
          tick pos=left,
          x grid style={darkgray176},
          xmajorgrids,
          xmin=-0.38, xmax=1.38,
          xtick style={color=black},
          xtick={0,1},
          xticklabels={FOM,ROM},
          y grid style={darkgray176},
          ymajorgrids,
          ymin=0, ymax=1.05,
          ytick style={color=black},
          ylabel = {relative simulation time $[-]$},
          ybar,
          legend image code/.code={\draw [#1,draw=none] (0cm,-0.1cm) rectangle (0.2cm,0.12cm); }
          ]
          \draw[draw=none,fill=steelblue31119180] (axis cs:-0.3,0) rectangle (axis cs:0.3,0.37625370829562);
          \addlegendimage{draw=none,fill=steelblue31119180}
          \addlegendentry{Assembly}

          \draw[draw=none,fill=steelblue31119180] (axis cs:0.7,0) rectangle (axis cs:1.3,0.36311980575586);
          \draw[draw=none,fill=darkorange25512714] (axis cs:-0.3,0.37625370829562) rectangle (axis cs:0.3,1);
          \addlegendimage{draw=none,fill=darkorange25512714}
          \addlegendentry{Solution}

          \draw[draw=none,fill=darkorange25512714] (axis cs:0.7,0.36311980575586) rectangle (axis cs:1.3,0.374717930988712);
    \end{groupplot}

\end{tikzpicture}
    \caption{Comparison of the solution time of a timestep, differentiating between the solution steps assembly and solution.  Normalized time the full order model (FOM) and the mortar reduced order model (ROM), for the "ring" example (left) and the "omega" example (right). }
    \label{fig:both_times}
\end{figure}

\subsubsection{Load-parametrization}
\label{sec:loadPara}

\begin{figure}[ht]
    \centering
    \def\svgwidth{0.7\textwidth}
\begingroup%
  \makeatletter%
  \providecommand\color[2][]{%
    \errmessage{(Inkscape) Color is used for the text in Inkscape, but the package 'color.sty' is not loaded}%
    \renewcommand\color[2][]{}%
  }%
  \providecommand\transparent[1]{%
    \errmessage{(Inkscape) Transparency is used (non-zero) for the text in Inkscape, but the package 'transparent.sty' is not loaded}%
    \renewcommand\transparent[1]{}%
  }%
  \providecommand\rotatebox[2]{#2}%
  \newcommand*\fsize{\dimexpr\f@size pt\relax}%
  \newcommand*\lineheight[1]{\fontsize{\fsize}{#1\fsize}\selectfont}%
  \ifx\svgwidth\undefined%
    \setlength{\unitlength}{575.99999135bp}%
    \ifx\svgscale\undefined%
      \relax%
    \else%
      \setlength{\unitlength}{\unitlength * \real{\svgscale}}%
    \fi%
  \else%
    \setlength{\unitlength}{\svgwidth}%
  \fi%
  \global\let\svgwidth\undefined%
  \global\let\svgscale\undefined%
  \makeatother%
  \begin{picture}(1,0.56250002)%
    \lineheight{1}%
    \setlength\tabcolsep{0pt}%
    \put(0,0){\includegraphics[width=\unitlength,page=1]{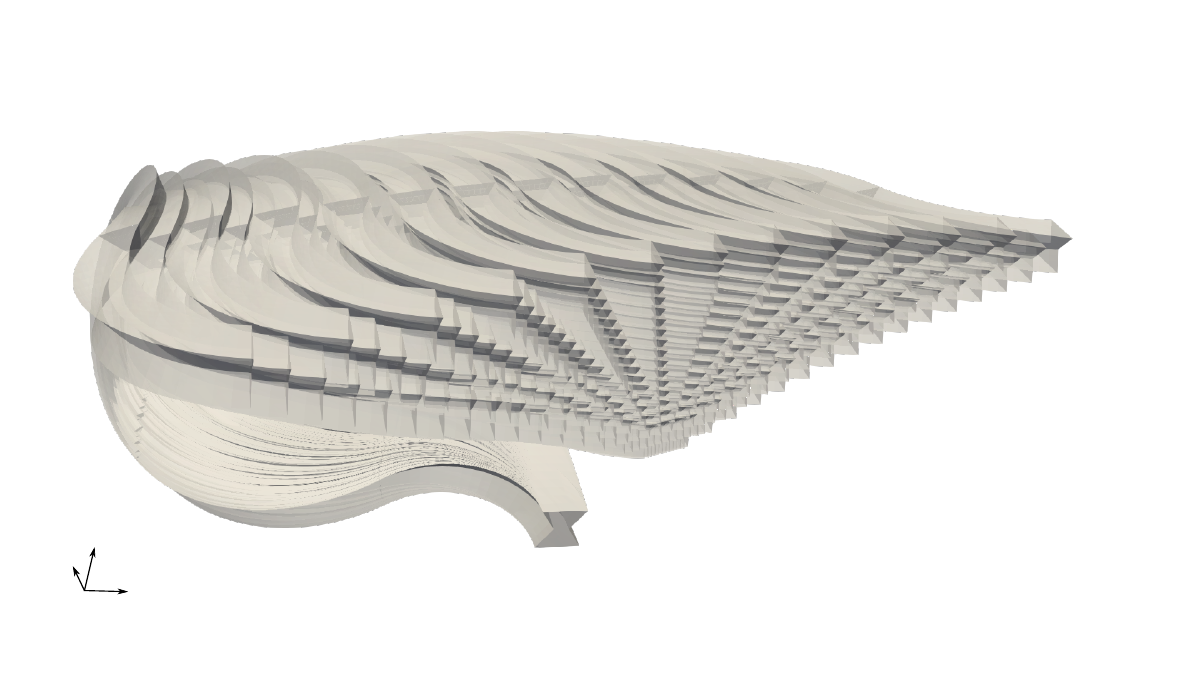}}%
    \put(0.03701247,0.07375706){\color[rgb]{0,0,0}\makebox(0,0)[lt]{\lineheight{1.25}\smash{\begin{tabular}[t]{l}$y$\end{tabular}}}}%
    \put(0.09281543,0.04819258){\color[rgb]{0,0,0}\makebox(0,0)[lt]{\lineheight{1.25}\smash{\begin{tabular}[t]{l}$x$\end{tabular}}}}%
    \put(0.06991842,0.11169415){\color[rgb]{0,0,0}\makebox(0,0)[lt]{\lineheight{1.25}\smash{\begin{tabular}[t]{l}$z$\end{tabular}}}}%
  \end{picture}%
\endgroup%

    \caption{All displacement states of the parametrized "omega" example. The $x$-displacements vary from $\bar u_x = -300\, \rm mm $ to $\bar u_x = 300\, \rm mm $. The displacement $\bar u_z = 150 \, \rm mm $ is constant. }
    \label{fig:alles_auf_einmal}
\end{figure}

This numerical example demonstrates the versatility of the modes and snapshots generated by the parametrization described in \Cref{sec:ring_snaps}. 
The geometry and boundary conditions are the same as in the "omega" example. 
The only difference is that we parametrize the applied displacement in $x$-direction. 
The $x$-displacements vary from $\bar u_x = -300\, \rm mm $ to $\bar u_x = 300\, \rm mm $.
The simulations are computed in increments of $50 \, \rm mm$. 
The displacement in $z$-direction is in all cases $\bar u_z = 150 \, \rm mm $. 
All displacement states are visualized in \Cref{fig:alles_auf_einmal}.

\Cref{fig:ux_parameter} shows the displacement error over the applied displacement $\bar u_x$. 
It can be seen, that for 80 modes per substructure the results are less accurate than the results computed with 100 or 120 modes per substructure. 
The solution with 120 modes per substructure is only at the ends of the parameter space more accurate than the solution with 100 modes per substructure. 
Between $\bar u_x = -100 \, \rm mm $ and $\bar u_x = 100 \, \rm mm $ 100 modes per substructure  give the most accurate results. 
A reason could be that the displacement state for this parameter range is better covered by the snapshots. 
For larger displacements the error can than be reduced by taking more modes into account. 

\begin{figure}[htbp] 
    \centering
\begin{tikzpicture}

\definecolor{darkgray176}{RGB}{176,176,176}
\definecolor{darkorange25512714}{RGB}{255,127,14}
\definecolor{forestgreen4416044}{RGB}{44,160,44}
\definecolor{lightgray204}{RGB}{204,204,204}
\definecolor{steelblue31119180}{RGB}{31,119,180}
\definecolor{crimson2143940}{RGB}{214,39,40}

\begin{axis}[
legend cell align={left},
legend style={font=\small,
              at={(0.8,0.97)},
              anchor=north east
            },
tick align=inside,
x grid style={darkgray176},
xlabel={$\bar u_x \: [\rm mm] $},
xmajorgrids,
xmin=-330, xmax=330,
ylabel={Error $ e \:[-]   $},
ymajorgrids,
ymin=0.000593177429614686, ymax=0.0070609548030892,
grid style = dashed, 
width=7.5 cm,
height=5.3 cm 
]
\addplot [very thick, darkorange25512714]
table {%
-300 0.00676696492247672
-250 0.00570929339990781
-200 0.00371168888342103
-150 0.00226746450775089
-100 0.00192331053400569
-50 0.00249317948106577
0 0.00291646445316763
50 0.00283881237058446
100 0.00280279148532142
150 0.00320023387536451
200 0.0039106996387357
250 0.0047233390183486
300 0.00551031730955809
};
\addlegendentry{$5\cdot80$}
\addplot [very thick, forestgreen4416044]
table {%
-300 0.00341222693423125
-250 0.00330220703651749
-200 0.00212725673456269
-150 0.00116023954348724
-100 0.00123152818496922
-50 0.00164885625194672
0 0.00181148446933883
50 0.00157673756865061
100 0.00125213285732127
150 0.00113618648439669
200 0.0012656742086748
250 0.00152574435866434
300 0.00184909631399998
};
\addlegendentry{$5\cdot100$}
\addplot [very thick, crimson2143940]
table {%
-300 0.00238193255632869
-250 0.00186710275177888
-200 0.00169102409384483
-150 0.00174167368784608
-100 0.00188794726264975
-50 0.00209952543863709
0 0.00213116516230986
50 0.00177485046582833
100 0.00124676355975532
150 0.000887167310227164
200 0.000891671413758841
250 0.0011048656609827
300 0.0013150281359908
};
\addlegendentry{$5\cdot120$}
\end{axis}

\end{tikzpicture}
    \caption{Plot of the displacement error norm over the applied displacement $\bar u_x$ for 80, 100, and 120 modes per substructure.}
    \label{fig:ux_parameter}
\end{figure}

\subsubsection{Viscoelasticity}\label{sec:Schlange_visco}  
In this section, we show that for inelastic material behavior, we can use the same modes as above. 
The modes are computed with a Neo-Hookean material (cf. \Cref*{sec:ring_snaps}) and are used now to reduce finite strain viscoelasticity. 
We simulated the "omega" boundary value problem now with a viscoelastic material law, according to \cite{reese_theory_1998,holthusen_inelastic_2023}. 
The displacement is applied over different time periods. 
The force-displacement curves in \Cref{fig:omega_force_disp_visco} show the rate dependence of the reaction forces. 
The slower the displacement is applied, the smaller the reaction force. 

The reduced computation is computed with 100 modes per substructure. 
The mode matrix is the same that we also used for the example with a Neo-Hookean material. 
It can be seen that the nonlinear force-displacement curves can be qualitatively reproduced by the reduced computation. 
For all three displacement rates the maximum force error is 
$(F_{\rm MOR} - F_{\rm Ref})/ F_{\rm Ref} \approx 0.04 \pm 0.001$. 
This error is similar to the errors of the elastic simulations \Cref{sec:Schlange_neoHooke}, where the maximum reaction force error was  $(F_{\rm MOR} - F_{\rm Ref})/ F_{\rm Ref} = 0.035$.
It can be concluded, that the elastic substructure modes can also be used to successfully predict the force response of viscoelastic simulations.

\begin{figure}[htbp]
    \centering
\begin{tikzpicture}

\definecolor{darkgray176}{RGB}{176,176,176}
\definecolor{darkorange25512714}{RGB}{255,127,14}
\definecolor{forestgreen4416044}{RGB}{44,160,44}
\definecolor{lightgray204}{RGB}{204,204,204}
\definecolor{steelblue31119180}{RGB}{31,119,180}

\begin{axis}[
legend cell align={left},
legend style={
  font=\small,
  at={(1.03,1.0)},
  anchor=north west
},
tick align=inside,
x grid style={darkgray176},
xtick style={color=black},
xlabel={displacement  [mm]},
xmin=-16.7705098312484, xmax=352.180706456217,
y grid style={darkgray176},
ylabel={Force $F$ [kN]},
ytick style={color=black},
ymin=-6.9247442938077, ymax=145.419630169962,
xmajorgrids = true, 
ymajorgrids = true, 
grid style = dashed,
width=8 cm,
height=6.2 cm 
]
\addplot [very thick,dashed, steelblue31119180]
table {%
0 0
20.9631372890605 20.0651151288161
41.9262745781211 40.1240147541849
62.8894118671816 59.2524082038922
83.8525491562421 77.5790530826446
104.815686445303 94.6357923752893
125.778823734363 109.598362576457
146.741961023424 121.385540519326
167.705098312484 129.119701819636
188.668235601545 132.78571133145
209.631372890605 133.412298463124
230.594510179666 132.534323744999
251.557647468726 131.556629377652
272.520784757787 131.471139572875
293.483922046847 132.852805142434
314.447059335908 135.959869347572
335.410196624968 140.851605818334
};
\addlegendentry{$t=1.6\, s$ ROM}
\addplot [very thick, steelblue31119180]
table {%
0 0
20.9631372890605 20.3471977462171
41.9262745781211 40.0653747947147
62.8894118671816 59.1609403588218
83.8525491562421 77.4476055045347
104.815686445303 94.450741390489
125.778823734363 109.333471463948
146.741961023424 120.994288558187
167.705098312484 128.528358274832
188.668235601545 131.899686597223
209.631372890605 132.133711946079
230.594510179666 130.771013159483
251.557647468726 129.217620445708
272.520784757787 128.463339026548
293.483922046847 129.088182848233
314.447059335908 131.366776214435
335.410196624968 135.379888884186
};
\addlegendentry{$t=1.6\, $s FOM}
\addplot [very thick,dashed, darkorange25512714]
table {%
0 0
20.9631372890605 16.095967342133
41.9262745781211 29.7493152382625
62.8894118671816 41.3260134049191
83.8525491562421 51.6841373555345
104.815686445303 60.9913414425009
125.778823734363 69.0505451827625
146.741961023424 75.4136924016403
167.705098312484 79.6440017728715
188.668235601545 81.6742874582708
209.631372890605 81.9819805527755
230.594510179666 81.4019631988863
251.557647468726 80.7804208363742
272.520784757787 80.7481138920327
293.483922046847 81.669371676841
314.447059335908 83.6938935292369
335.410196624968 86.8403807953271
};
\addlegendentry{$t=8 \,$s ROM}
\addplot [very thick, darkorange25512714]
table {%
0 0
20.9631372890605 16.2926401327156
41.9262745781211 29.6508430651804
62.8894118671816 41.2024949145165
83.8525491562421 51.5377271316091
104.815686445303 60.8160820357355
125.778823734363 68.8294888928807
146.741961023424 75.1161958671141
167.705098312484 79.2223953034276
188.668235601545 81.0666879177149
209.631372890605 81.1234862829676
230.594510179666 80.2331543855706
251.557647468726 79.2477484579615
272.520784757787 78.80197496381
293.483922046847 79.265980139978
314.447059335908 80.7988059416867
335.410196624968 83.4286737833603
};
\addlegendentry{$t=8 \, $s FOM}
\addplot [very thick,dashed, forestgreen4416044]
table {%
0 0
20.9631372890605 13.5603840336724
41.9262745781211 24.243566256688
62.8894118671816 33.3012570372916
83.8525491562421 41.7181281881196
104.815686445303 49.6403563855106
125.778823734363 56.8136481944742
146.741961023424 62.7338296561293
167.705098312484 66.892514792365
188.668235601545 69.1197140901582
209.631372890605 69.7757944950843
230.594510179666 69.5901497717643
251.557647468726 69.3244178197325
272.520784757787 69.5470051367974
293.483922046847 70.5838094988663
314.447059335908 72.5695725026029
335.410196624968 75.5263407990022
};
\addlegendentry{$t=16\,$s ROM}
\addplot [very thick, forestgreen4416044]
table {%
0 0
20.9631372890605 13.7024927021759
41.9262745781211 24.1439150124307
62.8894118671816 33.1927852674449
83.8525491562421 41.6014086709065
104.815686445303 49.5076504069649
125.778823734363 56.6483864414654
146.741961023424 62.5087519563103
167.705098312484 66.566365791352
188.668235601545 68.6394178736751
209.631372890605 69.086529868244
230.594510179666 68.64280166666
251.557647468726 68.0745692087887
272.520784757787 67.951545605656
293.483922046847 68.6019710736764
314.447059335908 70.1665017036602
335.410196624968 72.6742314198842
};
\addlegendentry{$t=16\, $s FOM}
\end{axis}

\end{tikzpicture}
    \caption{Force-displacement curves for different displacement rates considering finite strain viscoelasticity. The mortar approach is used for the reduced computation, with 100 modes per substructure.} 
    \label{fig:omega_force_disp_visco}
\end{figure}

\section{Discussion, conclusions and outlook}\label{sec:conclusion} 
In this paper, we have developed a substructuring technique using component-wise model order reduction and a mortar tied-contact formulation. 
In the numerical examples, we demonstrated that the developed method can predict solutions of systems constructed from reduced substructures. 
The method can handle non-matching meshes, different stiffnesses, geometric nonlinearities and material nonlinearities (finite strain viscoelasticity). 
The POD modes used for the substructure reduction were computed by simulating the substructures for different boundary conditions. 
We parametrized the boundary conditions on possible contact interfaces and used Latin hypercube sampling for the snapshot generation.   

We also compared the mortar approach, where we removed the Lagrange multipliers by static condensation, to the penalty approach used in \cite{zhou_proper_2018}. 
The mortar approach has multiple advantages compared to the penalty approach.  
The penalty approach has convergence problems for the here-discussed numerical examples. 
It also leads to ill-conditioned tangential stiffness matrices. 
We presume that the projection of this ill-conditioned tangential stiffness matrix leads to the observed convergence problems.
Not for all penalty parameters a solution can be found, and many more load steps are necessary compared to the mortar approach. 
The solution then also depends on the choice of the penalty parameter. 
The mortar approach does not have all these issues but has the disadvantage that the implementation is more complicated. 

The proposed method requires many modes per substructure to produce accurate results. 
In future works, local POD methods will be investigated to reduce the size of the POD bases of the substructures (cf. e.g.\cite{amsallem2012nonlinear,strazzullo_pod-based_2023}). 

The here proposed method still depends on the original dimensions of the problem. In the future, we will incorporate hyperreduction methods into the method. Possible methods could be energy conserving sampling and weighting (ECSW) (\cite{farhat_structure-preserving_2015}), or the discrete empirical interpolation method (DEIM) (\cite{chaturantabut_nonlinear_2010}). 
The method can then also be applied to other nonlinear mechanical substructuring problems, e.g. structural dynamics or mechanical meta-materials. 
In the future, we will also address other material nonlinearities, such as damage and plasticity (e.g. \cite{brepols2017gradient,kehls2023reduced})


\section{Declaration of competing interest} 
The authors declare that they have no known competing financial interests or personal relationships that could have appeared
to influence the work reported in this paper.

\section{Data availability} 
Data will be made available on request. 

\section{Acknowledgements}  
The authors gratefully acknowledge the funding granted by the German Research Foundation (DFG). 
The results presented here were developed within the subproject A01 of the Transregional Collaborative Research Center (CRC) Transregio (TRR) 280 with project number 417002380. 
Furthermore, T. Brepols, J. Kehls, and S. Reese gratefully acknowledge the funding that was granted within the subproject B05 "Coupling of intrusive and non-intrusive locally decomposed model order reduction techniques for rapid simulations of road systems" of the DFG CRC/TRR 339 with the project number 453596084, that was strongly involved in the origin of the paper.
The authors acknowledge the work of Hagen Holthusen whose element and material implementations where included into the model order reduction finite element program.

\section{Funding} 
German Research Foundation (DFG), project number 417002380

\appendix
\section{Appendix}
\label{sec:appendix}

\subsection{Stress Contour plots of Example 2}\label{apx:ring_omega} 

\begin{figure}[hptb]
    \centering
    \includegraphics[width=1\textwidth]{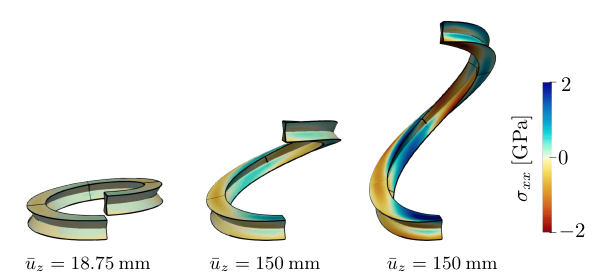}
    \caption{Shear stress contour plots of the deformed reduced solution of the "ring" boundary value problem. The system is solved by the substructuring MOR method with 80 modes per substructure. The black outline is the reference solution.} 
    \label{fig:contour_ring}
\end{figure}  

\begin{figure}[hptb]
    \centering
    \includegraphics[width=1\textwidth]{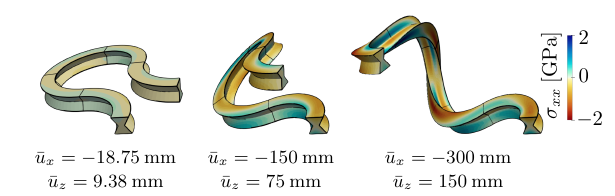}
    \caption{Shear stress contour plots of the deformed reduced solution of the "omega" boundary value problem. The system is solved by the substructuring MOR method with 120 modes per substructure. The black outline is the reference solution.} 
    \label{fig:contour_omega}
\end{figure}  


\bibliographystyle{agsm}
\bibliography{literature}

\end{document}